\documentclass{aa}  
\usepackage[varg]{txfonts}
\usepackage{natbib}
\usepackage{graphicx}
\usepackage{txfonts}
\usepackage{comment}

\newcommand{\oiii}{{[\ion{O}{iii}]}}
\newcommand{\neiv}{{[\ion{Ne}{iv}]}}

\newcommand{\nii}{[\ion{N}{ii}]}

\newcommand{\hii}{\ion{H}{ii}}

\newcommand{\ha}{H$\,\alpha$}
\newcommand{\hb}{H$\,\beta$}

\newcommand{\kms}{{$\rm{km\,s^{-1}}$}}
\newcommand{\degree}{{$^{\circ}$}}
\newcommand{\ergs}{$\rm{erg\,s^{-1}}$}
\newcommand{\msun}{$\rm{M_{\odot}}$}
\newcommand{\lsun}{$\rm{L_{\odot}}$}
\newcommand{\msunyr}{$\rm{M_{\odot}\,yr^{-1}}$}
\newcommand{\jykms}{$\rm{Jy\,km\,s^{-1}}$}
\newcommand{\mjykms}{$\rm{mJy\,km\,s^{-1}}$}
\newcommand{\Kkmspc}{$\rm{K\,km\,s^{-1}\,pc^{-2}}$}
\newcommand{\msunKkmspc}{$\rm{M_{\odot}\,[K\,km\,s^{-1}\,pc^{-2}]^{-1}}$}
\usepackage{siunitx}
\DeclareSIUnit{\angstrom}{\textup{\AA}} 

\newcommand{\ergscmAA}{$\rm{erg\,s^{-1}\,cm^{-2}\,\si{\angstrom}}$}

\newcommand{\vel}{{\rm{v}}}

\newcommand{\DL}{$D_{\rm{L}}$}
\newcommand{\sigmarms}{$\sigma_{rms}$}

\newcommand{\lagn}{$L_{\mathrm{AGN}}$}

\newcommand{\MBH}{$M_{\mathrm{BH}}$}
\newcommand{\Mstar}{$M_{\mathrm{*}}$}

\newcommand{\alphaCO}{$\alpha_{\rm{CO}}$}

\newcommand{\hst}{\textit{HST}}
\newcommand{\vla}{VLA}

\usepackage[dvipsnames]{xcolor}

\newcommand{\paperI}{Paper~I}

\defcitealias{dallagnol+25_paperII}{Paper~II}
\defcitealias{dallagnol+25_paperI}{Paper~I}

\usepackage{silence} 

\usepackage{hyperref} 
\hypersetup{
    colorlinks=true,
    citecolor=blue, 
    linkcolor=blue,
    filecolor=blue,      
    urlcolor=blue,
}

\usepackage{ulem} 
\usepackage{placeins}
\usepackage{amsmath}	
\usepackage{amssymb}	
\usepackage{subcaption}
\usepackage{newtxtext,newtxmath}

\begin{document}

   \title{NGC\,6860, Mrk\,915, and MCG\,-01-24-012}
   \subtitle{II. Inflowing and outflowing cold molecular gas and the connection with ionized gas in Seyfert galaxies}

    \titlerunning{NGC\,6860, Mrk\,915, and MCG\,-01-24-012. II. CO(2-1) dynamics}
    
    \authorrunning{Dall'Agnol de Oliveira et al.}

\author{Bruno Dall'Agnol de Oliveira\inst{\ref{inst1},\ref{inst2}}
   \and Thaisa Storchi-Bergmann\inst{\ref{inst2}}
   \and Neil Nagar\inst{\ref{inst3}}
   \and Santiago Garcia-Burillo\inst{\ref{inst4}}
   \and Rogemar A. Riffel\inst{\ref{inst5},\ref{inst5a}}
   \and Dominika Wylezalek\inst{\ref{inst1}}
   \and Pranav Kukreti\inst{\ref{inst1}}
   \and Venkatessh Ramakrishnan\inst{\ref{inst6}}
          }

   \institute{
Zentrum für Astronomie der Universität Heidelberg, Astronomisches Rechen-Institut, Mönchhofstr 12-14, D-69120 Heidelberg, Germany\label{inst1}\and
Departamento de Astronomia, Universidade Federal do Rio Grande do Sul, IF, CP 15051, 91501-970 Porto Alegre, RS, Brazil\label{inst2}\and
Astronomy Department, Universidad de Concepci\'on, Barrio Universitario S/N, Concepci\'on 4030000, Chile\label{inst3}\and
Observatorio Astronómico Nacional (OAN-IGN)-Observatorio de Madrid, Alfonso XII, 3, 28014 Madrid, Spain\label{inst4}\and
Departamento de F\'isica, CCNE, Universidade Federal de Santa Maria, 97105-900, Santa Maria, RS, Brazil\label{inst5}\and
Centro de Astrobiología (CAB), CSIC-INTA, Ctra. de Ajalvir km 4, Torrejón de Ardoz, 28850, Madrid, Spain\label{inst5a}\and
Finnish Centre for Astronomy with ESO, University of Turku, 20014 Turku, Finland\label{inst6}
   }

   \date{Received: / accepted: }

 
  \abstract
  {
We present a study of the cold molecular gas kinematics in the inner $\sim$\,4\,--\,7\,kpc (projected sizes) of three nearby Seyfert galaxies, with AGN luminosities of $\sim$\,10$^{44}$\,{\ergs}, using observations of the CO(2–1) emission line, obtained with the Atacama Large Millimeter/submillimeter Array (ALMA) at  $\sim$\,0.5\,--\,0.8{\arcsec} ($\sim$\,150\,--\,400\,pc) spatial resolutions.
After modeling the CO profiles with multiple Gaussian components, we detected regions with double-peak profiles that exhibit kinematics distinct from the dominant rotational motion. 
\\
In NGC\,6860, a molecular outflow surrounding the bipolar emission of the {\oiii} ionized gas is observed extending up to $R_{\rm{out}}$\,$\sim$\,560\,pc from the nucleus. There is evidence of molecular inflows along the stellar bar, although an alternative scenario, involving a decoupled rotation in a circumnuclear disk (CND) can also explain the observed kinematics.
\\
Mrk\,915 shows double-peak CO profiles along one of its spiral arms. 
Due to its ambiguous disk orientation, part of the CO emission can be interpreted as a molecular gas inflow or an outflow reaching $R_{\rm{out}}$\,$\sim$\,2.8\,kpc. \\
MCG\,-01-24-012 has double-peak profiles associated with a CND,  perpendicular to the {\oiii} bipolar emission.
The CO in the CND is rotating while outflowing within $R_{\rm{out}}$\,$\sim$\,3\,kpc, with the disturbances possibly being caused by the passage of the ionized gas outflow.
\\
Overall, the mass inflow rates are larger than the accretion rate needed to produce the observed luminosities, suggesting that only a fraction of the inflowing gas ends up feeding the central black holes. Although we found signatures of AGN feedback on the cold molecular phase, the mass outflow rates of $\sim$\,0.09\,--\,3\,\msunyr indicate an overall weak impact at these AGN luminosities. 
Nonetheless, we may be witnessing the start of the depletion and ejection of the molecular gas reservoir that has accumulated over time. 
}

   \maketitle

\section{Introduction}\label{sec:intro}

\begin{figure*}
	\includegraphics[width=1\linewidth]{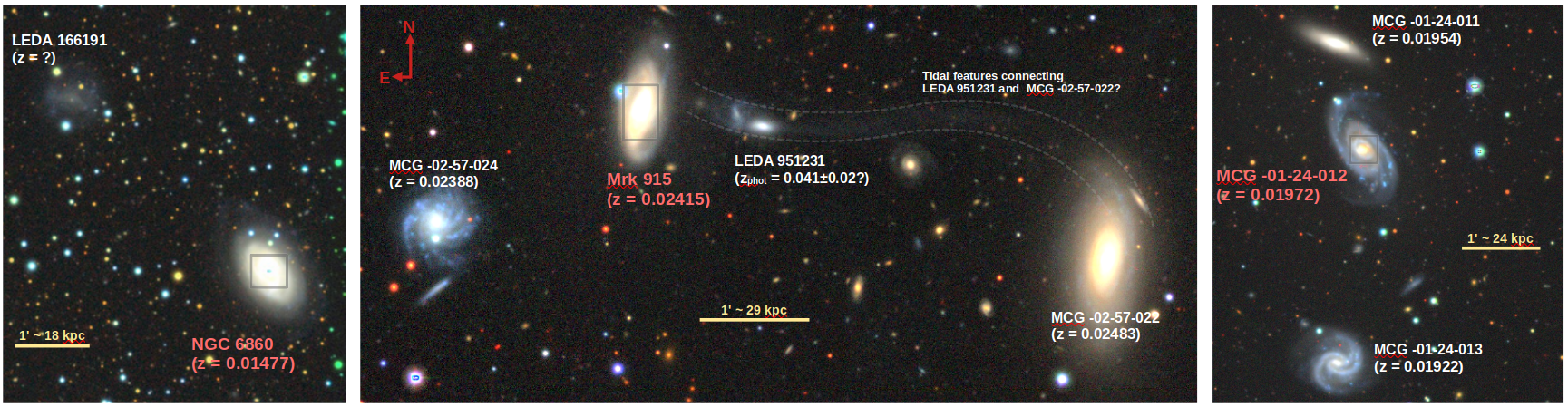}
    \caption{Color-composite images from DESI Legacy Survey Sky Viewer showing the local environment of NGC\,6860 (left), Mrk\,915 (middle) and MCG\,-01-24-012 (right), with squares marking the FoV of Fig.\,\ref{fig:moments_6860}, \ref{fig:moments_mrk915} and \ref{fig:moments_mcg0124102}. The spectroscopic redshifts of the brighter sources correspond to the preferred values from the NASA/IPAC Extragalactic Database (NED). 
    north is up and east is left in this and the remaining figures along the paper.} 
    \label{fig:companions}
\end{figure*}

Among the main physical processes that influence the evolution of galaxies are those occurring in Active Galactic Nuclei (AGN), ignited when matter is accreted into the supermassive black holes (SMBH) at the center of the host galaxies. 
Depending on the accretion rate, the energy can be released as radiation and winds from the accretion disk (radiative/quasar mode) or jets of highly energized particles (jet/mechanical mode) \citep{heckman_best14}.
How effectively this energy couples with the host galaxy's interstellar medium (ISM) is still a matter of debate. 
The net effect on the galaxy may lead to a suppression of the local star formation rate (SFR) in some objects \citep[negative feedback, e.g.][]{wylezalek+16,cicone+14}, as well as an increase in the SFR in others \citep[positive feedback, e.g.][]{gallagher+19,maiolino+17}.

One way to gauge the effect of feedback on the galaxy is to measure the mass outflow rate and its power \citep{harrison+18} and understand which aspects influence the accretion processes \citep{storchi_schnorr19}. 
Historically, most studies have used ionized gas to search for these feedback effects \citep[e.g.,][]{spence+18,dallagnol+21}.
However, the ionized phase constitutes only part of the gas in the ISM, which emphasizes the necessity of accounting for the impact on other gas phases \citep{cicone+18}.

An important phase to be studied is the cold molecular gas, with temperatures below $\sim$\,100\,K, which is the main ingredient for star formation \citep{veilleux+20}. 

If the AGN feedback disturbs kinematically the gas in this phase, the affected molecular content might not meet the physical condition required to form new stars, which can be viewed as a direct impact on the ISM. 
To quantify that, one can search for signs of disturbances in the cold molecular gas using emission lines that trace the total cold H$_2$ amount, such as the CO(2-1) molecular emission line \citep{bolatto+13}.

In the last two decades, evidence of AGN negative feedback in the cold molecular phase has emerged. 
In intermediate/luminous sources, with AGN bolometric luminosities of {\lagn}\,$\gtrsim$\,$10^{45}$\,{\ergs}, barely resolved observations indicate large mass outflow rates of \,$\sim$\,$10^2$\,--\,$10^3$\,{\msunyr} \citep[e.g.,][]{feruglio+10,cicone+14}. 
However, in studies using higher spatial resolution data (scales of 10\,--\,100 parsecs),  the measured effect appears to be lower, with values of $\sim$\,$10$\,--\,$10^2$\,{\msunyr} \citep[e.g.,][]{ramos-almeida+22,garcia-burillo+14}. 
For low/medium luminosity objects, {\lagn}\,$\lesssim$\,$10^{45}$\,{\ergs}, a wide range of resolved outflow rates of $\dot{M}_{\rm{mol,out}}$\,$\sim$\,0.1\,--\,10$^2$\,{\msunyr} have been reported \citep[e.g.,][]{garcia-burillo+14,oosterloo+17,audibert+19,alonso-herrero+19,slater+19,garcia-bernete+21,dallagnol+23,alonso-herrero+23}.

This work focuses on the low-luminosity regime by studying the CO(2-1) cold molecular gas kinematics of three nearby Seyfert galaxies: \object{NGC\,6860}, \object{Mrk\,915}, and \object{MCG\,-01-24-012}. The paper is organized as follows.
Sections\,\ref{sec:sample} and \ref{sec:sample_observations} describe the sample, the observations, and the archival data used.
The methodology used for the analysis is explained in Sect.\,\ref{sec:analysis}, with additional details in Appendices\,\ref{sec:grid}\,--\,\ref{ap:additional_fig}. 
We analyze and discuss each object individually in the Sects.\,\ref{sec:results_ngc6860} (NGC\,6860), \ref{sec:results_mrk915} (Mrk\,915)  and \ref{sec:results_mcg0124012} (MCG\,-01-24-012). %
The general discussion and conclusions are outlined in Sects.\,\ref{sec:Discussion} and \ref{sec:conclusions}.
Unless otherwise specified, all velocities are in the Kinematic Local Standard of Rest (LSRK). The luminosity distances and angular scales were calculated from the systemic redshift, for a $\mathrm{H_0=70\,km\,s^{-1}}$, $\mathrm{\Omega_M=0.3}$ and $\mathrm{\Omega_\Lambda=0.7}$ cosmology. 

\begin{table*}
	\centering
    \caption{
    General information about the sample. 
    } 
\begin{tabular}{ccccccccccc}
\hline\hline
    \# & Name &                  RA, DEC &       z &            \DL &                  Scale &          Type & log({\Mstar}) & log({\MBH}) & log({\lagn}) \\
      &         &    hh:mm:ss.ss dd:mm:ss.ss            &         & $\mathrm{Mpc}$ & $\mathrm{pc/\arcsec}$ & & $\mathrm{M_{\odot}}$ & $\mathrm{M_{\odot}}$ & $\mathrm{erg\,s^{-1}}$ \\
(1) & (2) & (3) & (4) & (5) & (6) & (7) & (8) & (9) & (10) \\
\hline
 1 &     NGC\,6860 & 20:08:46.89 -61:05:59.77 & 0.01477 &           64.0 &                  301 &     1.5 &    10.3 &     7.3\,--\,8.3 &  43.6\,--\,43.8 \\
 2 &      Mrk\,915 & 22:36:46.50 -12:32:42.80 & 0.02415 &            105 &                  487 & 1.5/1.9 &    10.0 &     7.3\,--\,8.4 &  44.1\,--\,44.3 \\
3 & MCG\,-01-24-012 & 09:20:46.26 -08:03:21.97 & 0.01972 &           85.7 &                  400 &   1.9/2 &   9.64 &      7.2$\pm$0.3 & 44.3\,--\,44.8 \\
\hline
\end{tabular}
\tablefoot{This is a reproduction of Table\,1 from \citetalias{dallagnol+25_paperI}, where the literature references for the last four columns are listed.
    (1) Source ID (for reference);
    (2) Galaxy name; 
    (3) RA and DEC coordinates of the ALMA millimeter continuum peak (in the ICRS frame); 
    (4) Redshift, corresponding to the systemic velocity (in the LSRK frame) of the disk model fitted to the data (see Sect.\,\ref{ap:bertola}); 
    (5) Luminosity distance; 
    (6) Angular scale; 
    (7) Seyfert type; 
    (8) AGN total luminosity; 
    (9) Stellar mass;  
    (10) Black Hole mass; 
    The range of values in {\MBH} and {\lagn} arises from the different measurements that we collected from the literature, with the exceptions calculated by us described below in the notes. 
    In particular for {\lagn}, the ranges are partly due to an observed intrinsic X-Ray variability.}
\label{tab:sample}
\end{table*}

\begin{figure*}[!ht]
    \centering
    \begin{subfigure}{.97\textwidth}
        \centering
    	\includegraphics[width=.99\linewidth]{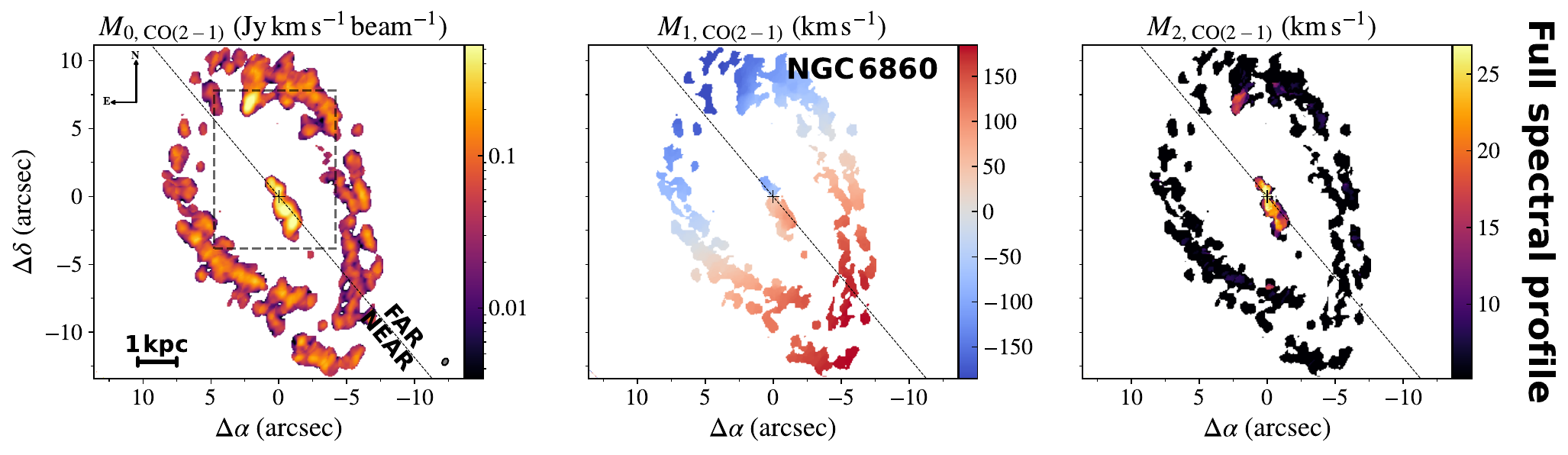}
        \caption{CO(2-1) moments of the observed spectral profile.}
    \label{fig:moments_6860}
    \end{subfigure}
    \begin{subfigure}{.97\textwidth}
        \centering
            \includegraphics[width=.99\linewidth]{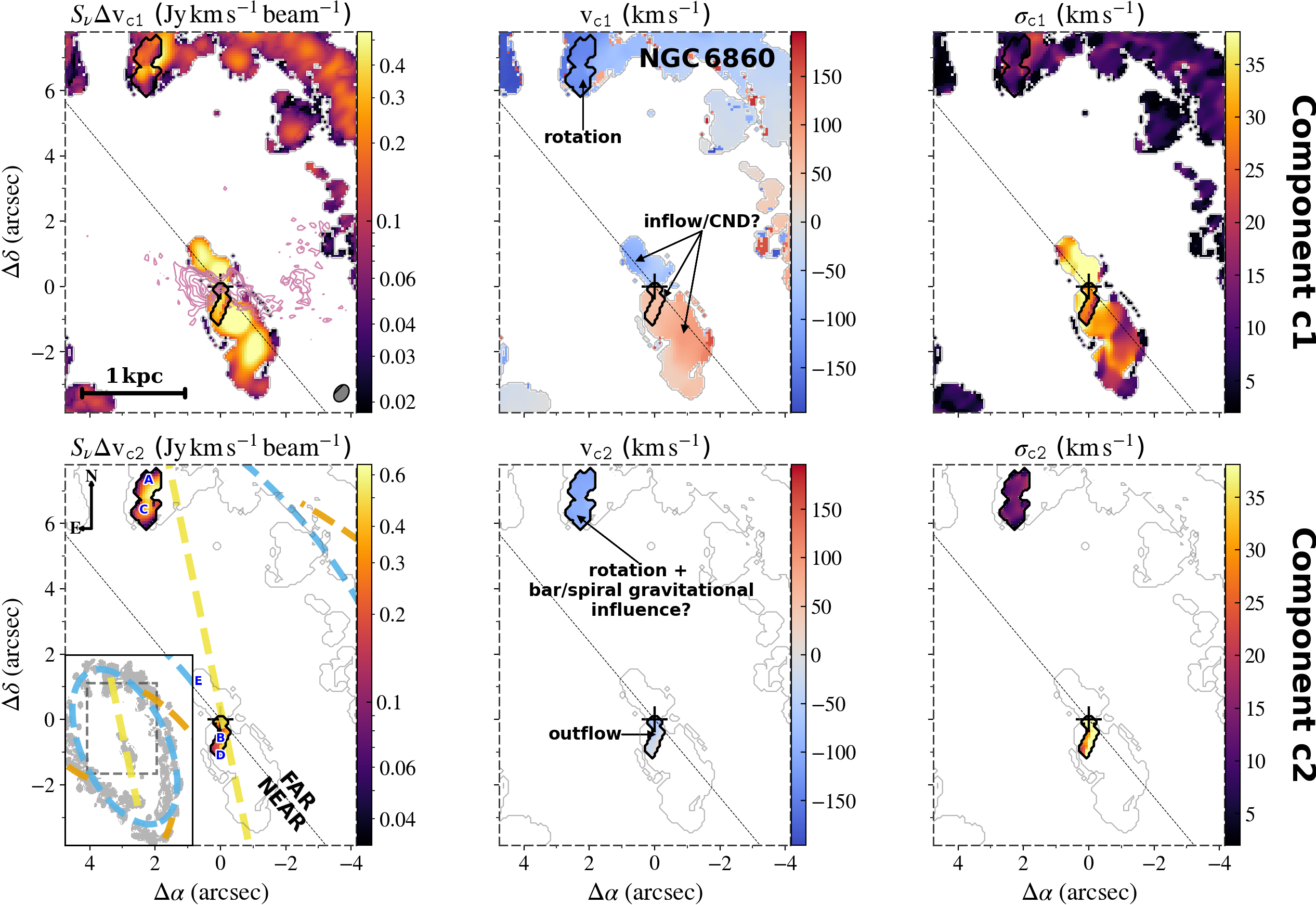} 
        \caption{Maps of the Gaussian parameters fitted to the CO(2-1).
                In the top left panel, there are ten reddish-purple contours representing the {\oiii} flux distribution, and ranging between 10$^{-18.3}$ and 10$^{-16.4}$\,{\ergscmAA}. Contours are evenly spaced logarithmically.
                }
    \label{fig:fitmaps_ngc6860}
    \end{subfigure}
\caption{
    Maps of NGC\,6860. 
    (a) Spatial distributions of the CO(2-1) moments' maps of the observed spectral profile, with $M_0$ corresponding to the integrated flux, $M_1$ tracing the mean radial velocity and $M_2$ the velocity dispersion of the CO profile.
    (b) Maps the parameters of the two Gaussian components fitted the CO(2-1) profiles: 
        \texttt{c1} (top row) and \texttt{c2} (bottom row), with the columns corresponding to distributions of the flux (left), the LoS velocity (middle), and the velocity dispersion (right). 
        In the bottom row, contours outline the full CO(2-1) extent (in gray) and the double-peak region (in black).  
        In the bottom left map, we added dashed lines representing the spiral arms (orange), the stellar bar (yellow), and the ring (sky-blue), as identified in \citetalias{dallagnol+25_paperI}.
        The letters A\,--\,E in the lower left panel show the locations of the spaxels used as examples of fits in Fig.\,\ref{fig:fitgrid}.
    For this and all other maps in the paper: north is up and east is left; coordinates are relative to ALMA millimeter continuum peak (black cross marks, see Table\,\ref{tab:sample}), assumed to be to the galactic nucleus; the gray ellipses correspond to the ALMA beam size of the observation; the inclined black dotted line is the major axis of the global kinematic model.
    }   
\label{fig:panel_ngc6860}
\end{figure*}

\begin{figure*}
        \centering
    \begin{subfigure}[b]{1\textwidth}
        \centering
    	\includegraphics[width=.98\linewidth]{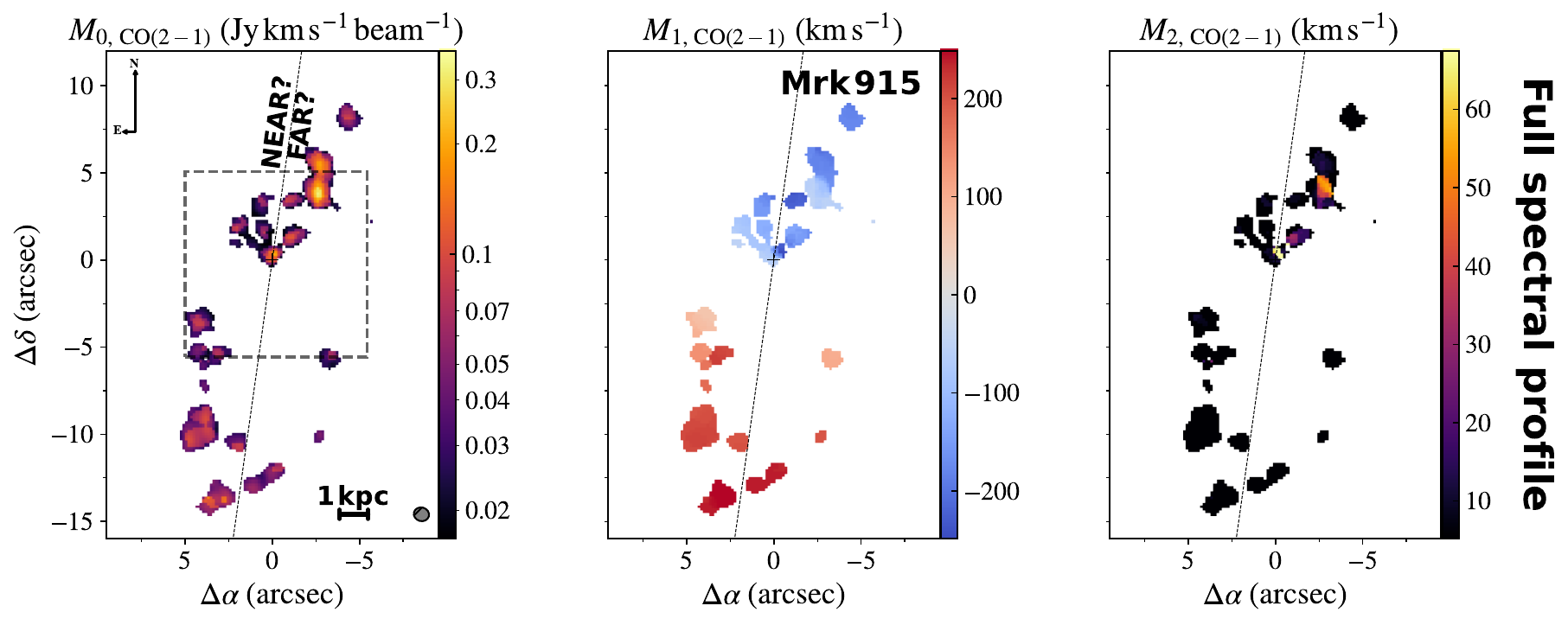}
            \caption{CO(2-1) moments of the observed spectral profile.}
            \label{fig:moments_mrk915}
    \end{subfigure}
    \begin{subfigure}[b]{1\textwidth}
        \centering
            \vspace{1ex}
            \includegraphics[width=.98\linewidth]{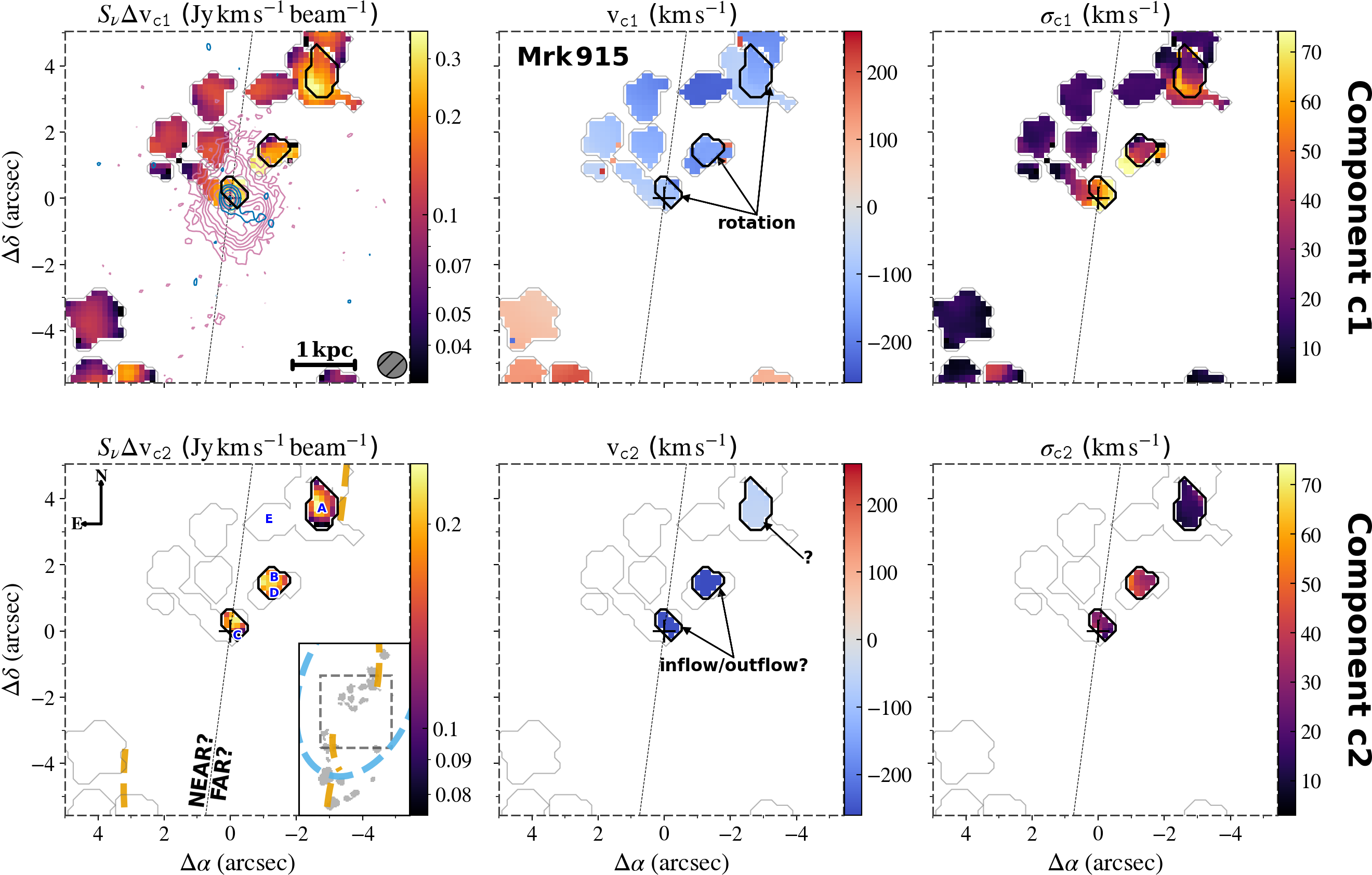} 
            \caption{Maps of the Gaussian parameters fitted to the CO(2-1). 
            In the top left panel, there are ten reddish-purple contour levels of {\oiii} flux distribution, 
            ranging between 10$^{-18.2}$ and 10$^{-15.7}$\,{\ergscmAA}. 
            The additional blue contours refer to {\vla} 3.6\,cm radio emission, and range between 0.01 and 1.20\,Jy.
            Contours are evenly spaced logarithmically.
            }
        \label{fig:fitmaps_mrk915}
    \end{subfigure}
    \caption{Same as Fig.\,\ref{fig:panel_ngc6860}, but for Mrk\,915.}
    \label{fig:panel_mrk915}
\end{figure*}

\begin{figure*}
        \centering
    \begin{subfigure}[b]{1\textwidth}
        \centering
    	\includegraphics[width=1\linewidth]{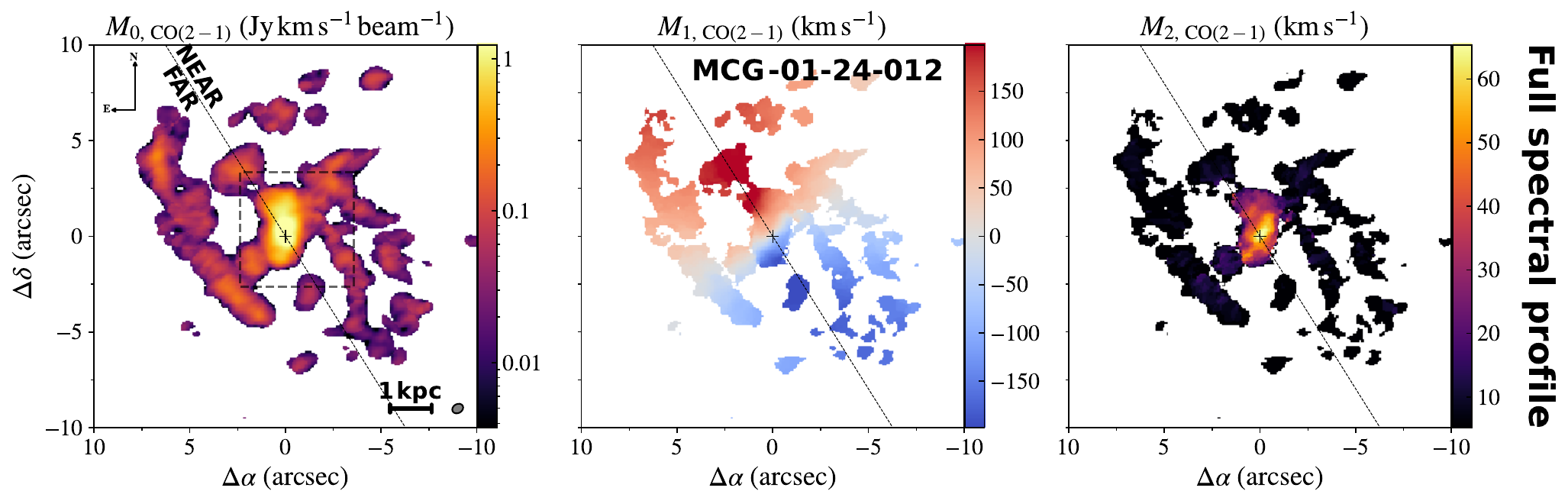}
            \caption{CO(2-1) moments of the observed spectral profile.}
            \label{fig:moments_mcg0124102}
    \end{subfigure}
    \begin{subfigure}[b]{1\textwidth}
        \centering
            \vspace{1ex}
            \includegraphics[width=1\linewidth]{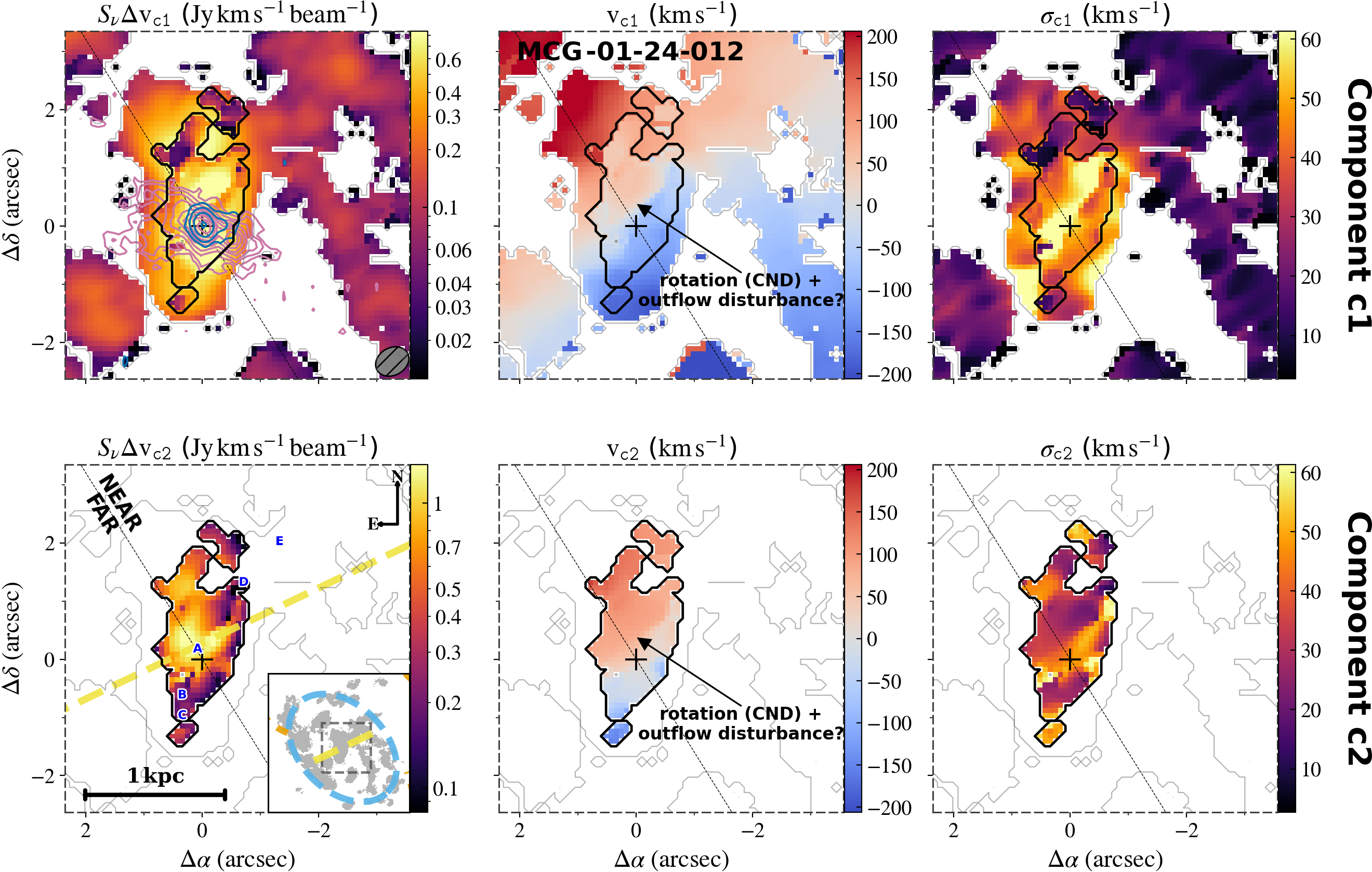} 
            \caption{Maps of the Gaussian parameters fitted to the CO(2-1). 
            In the top left panel, there are ten reddish-purple contour levels of {\oiii} flux distribution, ranging between 
            10$^{-18.3}$ and 10$^{-16.2}$\,{\ergscmAA}. 
            The additional blue contours refer to {\vla} 3.6\,cm radio emission, and range between 
            0.01 and 0.71\,Jy.
            Contours are evenly spaced logarithmically.}
            \label{fig:fitmaps_mcg0124012}
        \end{subfigure}
     \caption{Same as Fig.\,\ref{fig:panel_ngc6860}, but for MCG\,-01-24-012.}
    \label{fig:panel_mcg0124012}
\end{figure*}

\section{Sample}\label{sec:sample}

The three Seyfert galaxies studied here, \object{NGC\,6860}, \object{Mrk\,915} and \object{MCG\,-01-24-012}, have AGN bolometric luminosities  in the range of {\lagn}\,$\sim$\,10$^{43.6}$\,--\,10$^{44.8}$\,{\ergs}, with redshifts of 0.014\,$\lesssim$\,z\,$\lesssim$\,0.025. These are galaxies with prominent stellar structures, such as spiral arms, rings, and bars.
Their hosts have stellar masses in the range of {\Mstar}\,$\sim$\,10$^{9.6}$\,--\,10$^{10.3}$\,{\msun}, with black hole masses of {\MBH}\,$\sim$\,10$^{7.1}$\,--\,10$^{8.4}$\,{\msun} (see Table\,\ref{tab:sample}). A literature review and specific information about each object are provided below. The three galaxies are presented in color-composite images in Fig.\,\ref{fig:companions}, where we can observe their local environment.

The relation between the CO(2-1) cold molecular versus {\oiii}$\lambda\lambda$4959,5007 ionized gas flux distributions of these objects were presented in \citet[hereafter \paperI]{dallagnol+25_paperI}. The analysis revealed a spatial anticorrelation between the two gas phases, suggesting local CO depletion caused by AGN feedback.

\subsection{NGC\,6860}\label{sec:sample_ngc6860}
NGC\,6860 is a barred spiral (R')SB(r)b galaxy with both an inner and an outer stellar ring \citep{deVaucouleurs+91}. The nuclear spectrum was classified as Seyfert 1.5 by \citet{lipari+93}. 
Nonetheless, \citet{winter_mushotzky10} raised the possibility of it being a changing-look AGN  \citep{matt+03}, given the variability in X-ray flux and shape relative to previous observations \citep{winter+08}. 
Variations in the shape of the UV continuum and the {\ha} broad line region (BLR) profile were also observed in the past \citep{lipari+93}. 

There is no bright companion near NGC\,6860 (left of Fig.\,\ref{fig:companions}). The larger object at small angular distances is the unknown diffuse galaxy LEDA\,166191, separated by $\sim$\,3.5{\arcmin}. The nearest object with a similar redshift is 2MFGC\,15362 (z = 0.01495), located at $\sim$\,15{\arcmin} to the southeast of NGC\,6860.

\citet{lipari+93} presented a detailed analysis for this luminous IR source, using long-slit observations and narrow-band images centered on {\oiii}$\lambda\lambda$4959,5007 (hereafter, just \oiii) and {\ha}+{\nii}$\lambda\lambda$6548,6583. 
In the inner stellar ring, the {\oiii}/({\ha}+{\nii}) excitation map shows values in agreement with stellar-like ionization. 
Inside the ring, the above line ratio is  AGN-like, with the nuclear spectrum showing a broad H$\alpha$ component, denoting the presence of a BLR. The optical continuum shows old stellar population features in the spectra from regions along the bar, between the ring and the nucleus. In the ring, the spectra are more typical of {\hii} regions. 

\subsection{Mrk\,915}\label{sec:sample_mrk915}
Mrk\,915 is a spiral Sa galaxy \citep[][]{malkan+98}, possessing a variable Seyfert nucleus, that has been classified as type 1.5\,--\,1.9, depending on the presence of {\ha} and/or {\hb} BLR components \citep{goodrich1995,bennert+06,trippe+10}. 

Mrk\,915 is part of a triple system, along with other two bright sources \citep{karachentseva_karachentsev00}: MCG\,-02-57-024 and MCG\,-02-57-022, separated by angular distances of $\sim$\,2{\arcmin} and 4.3{\arcmin}, respectively (middle of Fig.\,\ref{fig:companions}). 
Moreover, there is another source (LEDA 951231) $\sim$\,1{\arcmin} to the west of Mrk\,915. It has long tidal features appearing to extend up to MCG\,-02-57-022 at southwest. We only found a highly uncertain photometric redshift for this object: $z_{\rm{phot}}$\,=\,0.041\,$\pm$\,0.02 \citep[DESI-DR9]{zhou+23}. Hence, we cannot confirm if this object is part of the system.

\citet{munoz-marin+09} presented {\hst} imaging observations in the Near Ultraviolet (NUV) for Mrk\,915. The extended NUV flux distribution is probably due to the \neiv$\lambda\lambda$3346,3426 emission lines since it is equivalent to the scaled {\oiii} distribution, although an NUV excess is observed at the nucleus \citep{munoz-marin+09}. 
Therefore, in addition to nuclear activity, a nuclear starburst could also contribute to the NUV excess. However, in this region, the gas line ratios indicate an AGN-like excitation \citep{bennert+06}, with the optical continuum being consistent with an elliptical galaxy template, indicating a dominant old stellar population \citep{trippe+10}. 

Mrk\,915 has multiple X-ray observations  \citep{trippe+10,severgnini+15,serafinelli+20}. \citet{ballo+17} identified X-ray variability in intensity, but not in the shape of the continuum. Their models advocates for the presence of warm absorbers, but with the continuum variability being mostly caused by an intrinsic variation in nuclear power.

\subsection{MCG\,-01-24-012}\label{sec:sample_mcg0124012}
MCG\,-01-24-012 is a spiral SAB(rs)c galaxy showing both an inner stellar ring and weak bar \citep[][]{deVaucouleurs+91}. Its nucleus was originally classified as a type 2 AGN \citep[e.g.,][]{veron-cetty_veron06}, but a broad FWHM\,$\sim$\,2000\,{\kms} component was later detected in the Pa\,$\beta$ emission line \citep[e.g.,][]{onori+17}, as well as a weak broad component in the {\ha} line \citep{la_franca+15}. The ``hidden'' BLR component in the Balmer lines may be a consequence of the optical spectral region being more attenuated relative to the infrared by the material in the host galaxy \citep[e.g.,][]{ricci+22}. 

MCG\,-01-24-012 has two close and bright companion galaxies: MCG\,-01-24-011 and MCG\,-01-24-013, separated by angular distances of $\sim$\,1.3{\arcmin} and 2.7{\arcmin}, respectively (see (Fig.\,\ref{fig:companions})). Together, they appear to form a triple system, probably gravitationally bound due to the small differences in redshift.

The X-ray continuum of MCG\,-01-24-012 has been studied in the past \citep{malizia+02,winter+09}, showing moderate variability \citep{middei+21}. Besides the Fe\,K$\alpha$\,6.4\,keV fluorescent emission line, there is an absorption feature detected at $\sim$\,7.5\,KeV \citep{malizia+02}. If corresponding to the Fe\,XXVI\,Ly$\alpha$ line (rest 6.97\,keV), this feature could originate from a powerful nuclear outflow with velocity V$_{\rm{out}}\,{\sim}\,0.06\,c$ \citep{middei+21}.

\section{Observations}\label{sec:sample_observations}

We observed our sources using the \textit{Atacama Large Millimeter/submillimeter Array} (ALMA) during Cycle 6 (ID:
2018.1.00211.S), with one of the spectral windows (SPW) centered on the CO(2-1) emission line (230.538\,Hz rest frequency). 
The final continuum-subtracted cubes have channel widths of $\Delta \vel$\,$\sim$\,10.2\,{\kms}, with a {\sigmarms} noise at their field-of-view (FoV) centers ranging between $\sim$\,0.4 and 0.7$\,\mathrm{mJy\,beam^{-1}}$. 
The FWHM of the synthesized beams of NGC\,6860, Mrk\,915, and MCG-01-24-012 are, respectively, 0.41{\arcsec}$\times$\,0.56{\arcsec} ($\sim$\,150\,pc),  0.78{\arcsec}$\times$\,0.88{\arcsec} ($\sim$\,400\,pc) and  0.47{\arcsec}$\times$\,0.6{\arcsec} ($\sim$\,200\,pc). The size of the individual spaxels are 0.076, 0.16, and 0.087{\arcsec}, in the same order. 

We also used archival data from the \textit{Hubble Space Telescope} ({\hst}) and \textit{Very Large Array} ({\vla}). The {\hst} FR533N narrow-band images  \citep[ID: 8598,][]{schmitt+03}, centered on the {\oiii}$\lambda\lambda$4959,5007 emission lines (hereafter {\oiii}),  were used to trace the ionized gas flux distribution. The {\vla} 8.46\,GHz (3.54\,cm) radio images \citep[proposal ID: AA226]{schmitt+01}, was used to look for signatures of radio jets. 
We also collected archival DECam (Dark Energy Camera) g and z-bands images from the Data Release 10 of the DESI (Dark Energy Spectroscopic Instrument) Legacy Imaging Surveys.

Details about observations and reductions of the ALMA and the archival data are given in \citetalias{dallagnol+25_paperI}.

\section{Analysis}\label{sec:analysis}

\subsection{Moments from CO(2-1)}\label{sec:moments}

An overview of the spatial distribution and kinematics of the molecular gas can be obtained from the moments of the CO line profiles (as described in Appendix\,\ref{ap:mom-mask}). Figs.\,\ref{fig:moments_6860}, \ref{fig:moments_mrk915}, and \ref{fig:moments_mcg0124102} show the resulting maps of the moments in the first row: for NGC\,6860, Mrk\,915, and MCG\,-01-24-012, respectively.  
The 0-th moment ($M_0$) is the distribution of the line flux, the first moment ($M_1$) is the intensity-weighted mean of the line of sight (LoS) velocity, and the second moment ($M_2$) is the square root of the intensity-weighted mean of the squared velocity dispersion \citep[e.g.,][]{ramakrishnan+19}.
$M_1$ traces the projected centroid velocity of the CO clouds inside a given region, while $M_2$ traces the width of the profiles (sensitive to kinematic disturbances). 
For a line profile perfectly modeled by a single Gaussian, $M_1$ and $M_2$ are the centroid velocity and the velocity dispersion of the gas, respectively.

\subsection{Dust attenuation and orientation of the disk}\label{sec:dust}
The interpretation of the kinematics will depend on knowing which are the near and far sides of the galaxies' disks: the orientation relative to the plane of the sky. For this, we assumed that the side where the optical stellar continuum is more attenuated by dust is the near side. 
This is based on the overall assumption that,  close to the photometric minor axis (measured in optical broad-band images), the presence of dust lanes is more ``noticeable'' on the near side of the disk due to the contrast with the background stellar light from the bulge \citep{iye+19}. 
We also checked if the resulting orientation agrees with the galaxy having trailing spiral arms. 
If the molecular gas and stars co-rotate, the CO(2-1) first moment maps ($M_1$) can be used to gauge the sense of rotation of the spiral arms (middle column of Figs.\,\ref{fig:moments_6860}, \ref{fig:moments_mrk915} and \ref{fig:moments_mcg0124102}). For reference, the photometric major axes of NGC\,6860, Mrk\,915, and MCG\,-01-24-012 are 34, 162, and 38{\degree} (mean position angles), respectively \citep[from broad-band B images]{schmitt+00}.

We used the ratio between the DECam g and z-band images as a proxy for dust attenuation in the host galaxy, since the extinction is stronger at bluer wavelengths \citep[e.g.,][]{mezcua+15}. The results are shown in Fig.\,\ref{fig:gz}. 
Inspecting these maps, we identified that the near sides of MCG\,-01-24-012 and NGC\,6860 are to the southeast and northwest, respectively. These orientations agree with the galaxies having trailing arms. 
For Mrk\,915, the orientation is ambiguous: while the dust attenuation map indicates that the near side may be to the southeast, a trailing arms scenario implies that the near side of the disk is to the northwest. We discuss this scenario in Appendix\,\ref{ap:mrk915-orientation}. Both orientations for the Mrk\,915's disk were considered in the analysis done in Sect.\,\ref{sec:results_mrk915}.

\subsection{CO(2-1) emission-line fitting}\label{sec:line-fitting}

In all sources, CO(2-1) emission-line profiles show double peaks near the nucleus, typically within a radius of $\sim$\,1\,kpc, but also in a few locations far from the nucleus.
We tentatively attribute the origin of these two peaks as coming from different molecular gas clouds with distinct global kinematics. 
We modeled the CO profiles with multiple Gaussian components using the {\sc python} package  {\sc ifscube} \citep{ruschel-dutra+21}. For each spaxel (sizes of 0.08\,--\,0.16{\arcsec}\,$\sim$\,20\,--\,80\,pc), the program returns the Gaussian parameters from the model that minimizes the residuals. The optimization is done by using the Sequential Least Squares Programming (SLSQP) method from {\sc scipy.minimize}.  

To account for the presence of double peaks, each spectrum was fitted with two models: a single Gaussian (\texttt{1g}), and two Gaussian curves (\texttt{2g}). 
Each Gaussian component is characterized by its amplitude ($S_{\nu,\rm{CO}}$), line of sight velocity ($\vel_{\rm{CO}}$) and  velocity dispersion ($\sigma_{\rm{CO}}$).
The velocity-integrated flux ($S_{\nu} \Delta \vel_{\rm{CO}}$) in units of {\jykms} can be converted to luminosities in units of {\Kkmspc} ($L'_{\rm{CO}}$) and {\lsun} ($L_{\rm{CO}}$) by following \citet[Eqs. 1 and 3]{solomon+97}.

\begin{figure*}
	\includegraphics[width=1\linewidth]{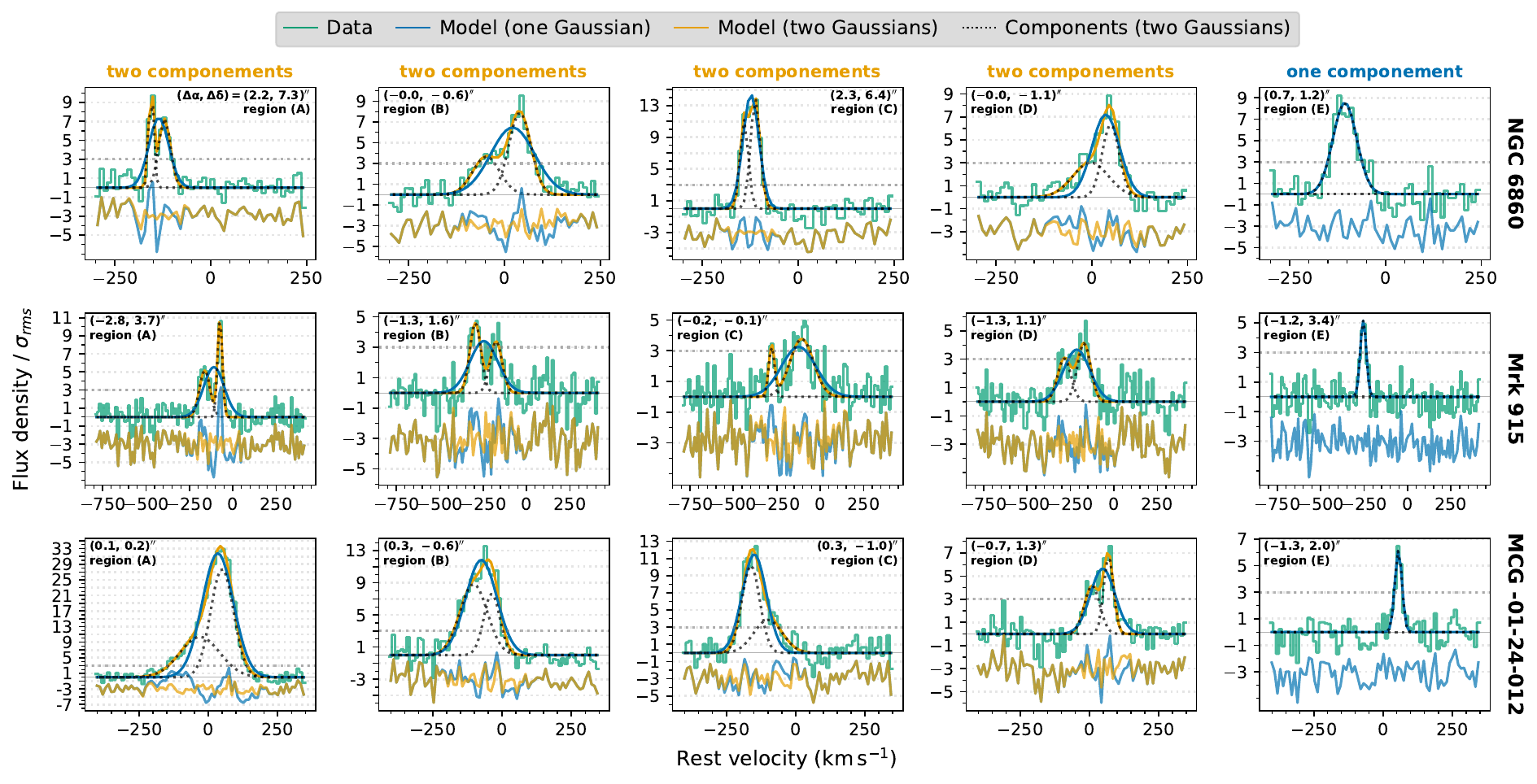}
    \caption{Example of the model multiple Gaussian models fitted to the CO(2-1) profiles of individual spaxels (0.076\,--\,0.16{\arcsec} sizes), one row per object.  The first four columns (A to D) show examples where two Gaussian components were preferred over a single component, as in the last column (E). 
    The spatial positions  (relative to the nucleus) of each spaxel are shown at the top of each panel, and are marked in Figs.\,\ref{fig:fitmaps_ngc6860}, \ref{fig:fitmaps_mrk915} and \ref{fig:fitmaps_mcg0124012} (blue letters in the lower left panels).
    The lines represent: 
    original data (bluish-green); 
    one Gaussian component models (in blue);
    two Gaussian models (in orange), along with their individual components (in black dashed).
    The residuals of both models are shifted downwards in the plots.
    The y-axes are in units of S/N, with the horizontal dotted line corresponding to S/N = 3 being highlighted, since this was the threshold used to discard components.
    }
    \label{fig:fitgrid}
\end{figure*}

Gaussian components with a signal-to-noise ratio (S/N) below 3 were discarded. Therefore, \texttt{2g} models were selected over \texttt{1g} ones when both Gaussian components (of the \texttt{2g} model) exceeded the above S/N criterion. 
Since neighbors' spaxels are non-independent inside the beam area, 
isolated spaxels with \texttt{2g} were also discarded.
Examples of CO profile fits are shown in Fig.\,\ref{fig:fitgrid}. 
Maps of the best-fit parameters are displayed in Figs.\,\ref{fig:fitmaps_ngc6860}, \,\ref{fig:fitmaps_mrk915}, and  \ref{fig:fitmaps_mcg0124012}, showing the distributions of the flux, LoS radial velocity, and velocity dispersion of each component. We also generated a grid of spectra in Fig.\,\ref{fig:grids} (Appendix\,\ref{sec:grid}), which allows a comparison of CO profiles inside and outside the identified double-peak regions.

\subsection{Position velocity diagrams}\label{sec:pv}
Figure\,\ref{fig:pv} shows the CO(2-1) position-velocity (PV) diagrams for  three orientations:  
(1) along the kinematic major axes (see Section\,\ref{sec:fit-bertola}): position angles (PA) of 40, $-7$, and 32{\degree}, for NGC\,6860, Mrk\,915, and MCG\,-01-24-012, respectively;
(2) along the minor axes, perpendicular to the above orientations; 
(3) along a third direction, covering most of the double-peak region:  position angles of 16, $-36$, and 0{\degree}, for NGC\,6860, Mrk\,915, and MCG\,-01-24-01.

\subsection{Kinematic modeling}\label{sec:fit-bertola}

In all sources, most of the CO profiles far from the nucleus show narrow single-component profiles. These regions in the $M_1$ moment maps (Fig.\,\ref{fig:moments_6860}, \ref{fig:moments_mrk915} and \ref{fig:moments_mcg0124102}) display a typical disk rotation pattern and low $M_2$ values ($\lesssim$\,15\,{\kms}), indicating that they might be tracing gas rotating in the galaxy disk.

To verify this, we fitted a 2D disk rotation model \citep{bertola+91} to the data. More specifically, we fitted the line of sight (LoS) velocity distribution map in regions showing only single components dominated by rotation: the LoS velocity map of the \texttt{c1} component ($\vel_{\tt{c1}}$), excluding the region with double-peak profiles (see detailed description in Appendix\,\ref{ap:bertola}). 

The best-fit parameters of the models are displayed in the first three rows of Table\,\ref{tab:bertola2}. 
Maps of the data, best-fit model, and residuals are shown in Fig.\,\ref{fig:bertola} (first three columns), with a zoom-in of the residuals from both components displayed in the last two columns. They are also represented by bluish-green solid lines in the PV diagrams of Fig.\,\ref{fig:pv}.

For NGC\,6860, the CO data in the bar were also masked during the fit, with only data in the ring being used for the global rotating disk modeling. This was done because the kinematics in the bar and the ring diverge noticeably, with a simple rotating model being unable to account for both dynamics.
We note that including the data from the bar in a global fit would not significantly change the results (see the fourth row in Table\,\ref{tab:bertola2}). The last three rows of the table refer to alternative models discussed in Sects.\,\ref{sec:ngc6860_res_bar_CND} and \ref{sec:mcg0124012_res_doublepeak}.

\begin{table*}
    \centering
    \caption{
    Parameters of the 2D disk models.
    }
\begin{tabular}{cccccccc}
\hline\hline
Name &                Data fitted               &   $\vel_{\rm{sys}}$ &                    PA &                   $i$ &     V$_{\rm{max}}$ &                        $c_0$ &  $p$ \\
  &  & $\mathrm{km\,s^{-1}}$ & $\mathrm{{}^{\circ}}$ & $\mathrm{{}^{\circ}}$ & $\mathrm{km\,s^{-1}}$ & $\mathrm{{}^{\prime\prime}}$ &      \\
 (1) &  (2) &  (3) &  (4) &  (5) & (6) & (7) & (8) \\
\hline
 \multicolumn{8}{c}{\it (global models)}\\
NGC\,6860 (only ring) & $\vel_{\tt{c1}}${$^a$}  & 4394.1$\pm$0.1  &  40.0$\pm$0.1   &  57.7$\pm$0.2  &  266$\pm$1     &  9.0$\pm$0.1  &  1     \\
Mrk\,915             & $\vel_{\tt{c1}}${$^a$}   & 7154.3$\pm$0.4  &  172.4$\pm$0.1  &  66.5$\pm$0.2  &  257.2$\pm$0.8   &  1.2$\pm$0.1  &  1    \\
MCG\,-01-24-012      & $\vel_{\tt{c1}}${$^a$}   & 5854.0$\pm$0.2  &  32.0$\pm$0.1   &  50.2$\pm$0.3  &  453$\pm$6    &  1.6$\pm$0.2  &  1.4$\pm$0.1  \\
\hline
  \multicolumn{8}{c}{\it (alternative models)}\\
%
NGC\,6860 (ring\,+\,bar)  &  $\vel_{\tt{c1}}${$^a$}                &  4393.7$\pm$0.1 & 39.1$\pm$0.1 &  56.4$\pm$0.2 & 233$\pm$1 &  6.0$\pm$0.1  & 1 \\
NGC\,6860 (only bar)   &  $\vel_{\tt{c1}}${$^a$}                   &  4394.1$^c$ & 39.8$\pm$0.3 &  69.3$\pm$0.5 &  153$\pm$6 & 1.0$\pm$0.1 & 1.3$\pm$0.1 \\
MCG\,-01-24-012 (CND)  & ($\vel_{\tt{c1}}$+$\vel_{\tt{c2}}$)/2$^b$ & 5854.0$^c$    &  35$\pm$1     &  3$\pm$1   &  2300$\pm$400  &  0.8$\pm$0.1  &  1 \\
MCG\,-01-24-012 (CND)  & ($\vel_{\tt{c1}}$+$\vel_{\tt{c2}}$)/2$^b$ & 5854.0$^c$    &  0$^c$    &  74$\pm$2  &  270$\pm$40  & 1.6$\pm$0.2  & 1.5$\pm$0.1   \\
\hline
\end{tabular}
\tablefoot{
The first three rows refer to the main fits, while the remaining rows refer to alternative versions. Parameters corresponding to 2D disk rotation models \citep{bertola+91}. 
    (1) Galaxy name;
    (2) Data used for the fit; 
    (3) Systemic relativistic velocity (in the LSRK frame); 
    (4) Position Angle; 
    (5) Inclination of the disk (relative to the plane of the sky); 
    (6) Maximum circular velocity; 
    (7) Spatial scale parameter; 
    (8) Profile shape parameter, with values without uncertainties having errors below 0.01.
    The kinematic center was fixed at the millimeter continuum peak (Table\,\ref{tab:sample}).
\\
$^{a}$: Using only data from the single-peak region. \par
$^{b}$: Using only data from the double-peak region. \par
$^{c}$: Parameter fixed during the fit. \par
}
\label{tab:bertola2}
\end{table*}

\subsubsection{Identification of the components in the double-peak region}\label{sec:identification components}

In the region with double-peak profiles, the origin of the emission of each component (leading to the double peaks) is not known a priori. As a first identification criterion, we assumed that one of them is tracing gas dominated by rotation. 
Therefore, we named the component with lower absolute residuals relative to the disk model as ``\texttt{c1}''. 
The other component was called ``\texttt{c2}'', and, in principle, traces the emission from disturbed gas (e.g., from outflows or inflows). 
Whenever needed, we added a second criterion for selecting \texttt{c1}: we visually selected it to be the one that minimizes the discontinuity in the \texttt{c1} maps. 
In some cases, none of the components follow the global disk rotation pattern (e.g., CO along the bar in NGC\,6860, as discussed in Sects.\,\ref{sec:ngc6860_res_bar_inflow} and \ref{sec:ngc6860_res_outflow}).
We emphasize, therefore, that the \texttt{c1} component will thus \textit{not be} solely tracing gas rotating in the galaxy disk.


\subsection{Schematic models}\label{sec:models}

\begin{figure*}
	\includegraphics[width=1\linewidth]{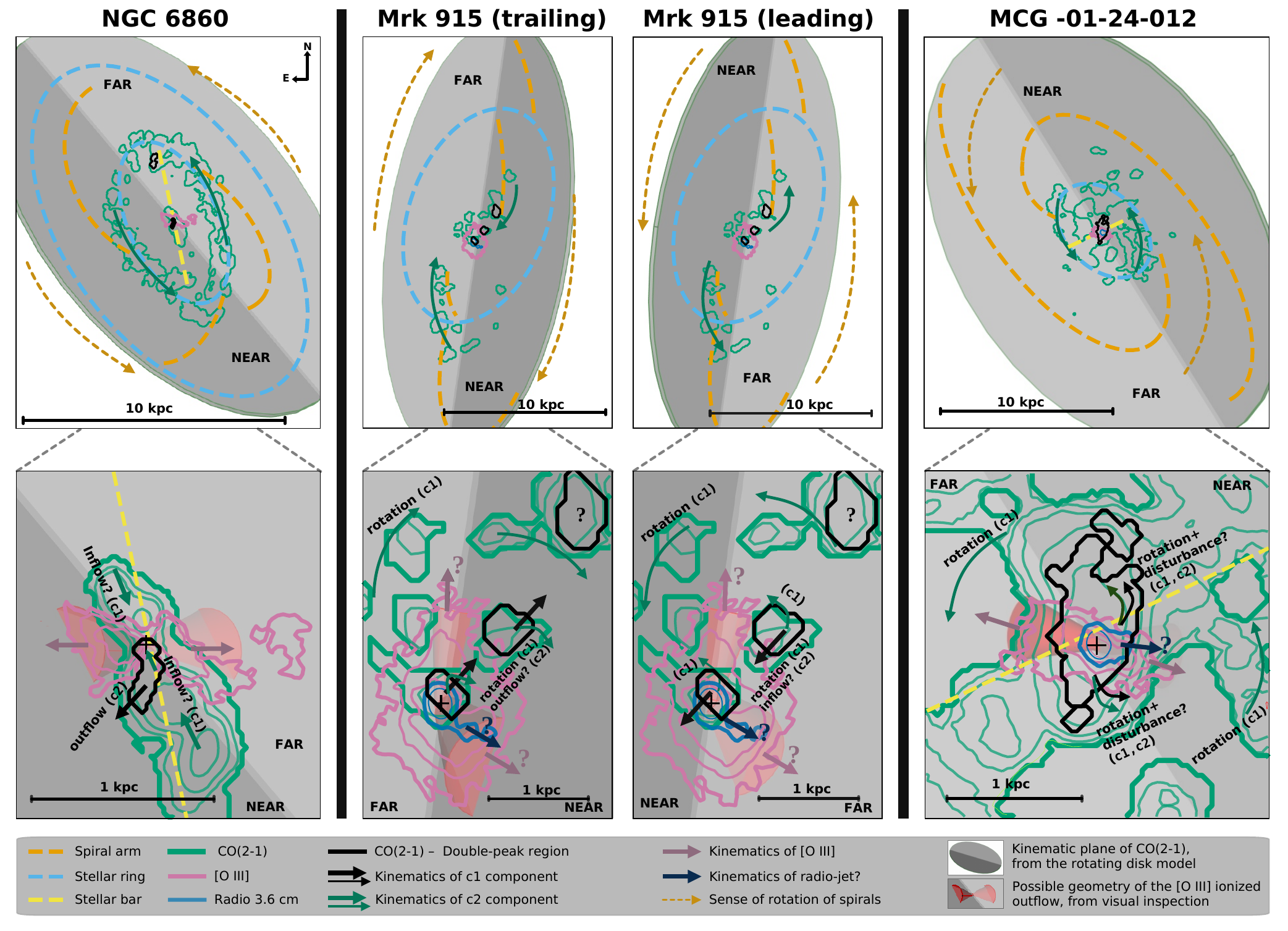}
    \caption{Schematic models for our proposed scenarios of NGC\,6860, Mrk\,915, and MCG\,-01-24-012, with the second row corresponding to a zoom-in in the central regions. 
    Due to the ambiguous Mrk\,915's disk orientation (see Appendix\,\ref{ap:mrk915-orientation}), we added both scenarios of spiral arms motion: trailing and leading. 
    The contours represent the flux distribution of: cold molecular gas from CO(2-1) $M_0$ (bluish-green), ionized gas from {\hst} {\oiii} (reddish-purple), VLA 3.6\,cm radio (blue), and the CO double-peak region (black).  For a better view, we plotted only a few contour levels and manually cleaned isolated ones. 
    Stellar structures are drawn in dashed lines: spiral arms (in orange), stellar rings (sky-blue), and bars (yellow).  
    The gray ``3D'' planes 
    correspond to the plane of the global rotating disk model. 
    We also added ``3D'' bipolar ionized cones to represent possible orientations of these structures. 
    The senses of rotation of the spirals are shown in orange dashed arrows, assuming that they ``follow'' rotation of the CO. 
    The solid arrows represent the direction of motion of: components \texttt{c1} (blueish-green) and \texttt{c2} (black) of the CO(2-1), ionized gas (reddish-purple) and radio-jets (blue), with curved lines representing rotation. 
    We added question marks to represent that we can only confirm the kinematics of the {\oiii} using resolved integral-field observations, and the same for the possible radio jets that need deeper and higher spatial resolution observations. 
    In the double-peak regions, we added one arrow for each component.  In particular, for MCG\,-01-24-012, we tried to represent that both components are dominated by rotation, with each one having an extra positive/negative radial velocity ``kick'' due to the CO being disturbed equatorially by the ionized outflow. 
    }
    \label{fig:models}
\end{figure*}

Figure\,\ref{fig:models} displays schematic models for the three objects. They will help visualize the scenarios discussed in Sects.\,\ref{sec:results_ngc6860}\,--\,\ref{sec:results_mcg0124012}. 
See the upper left panels of Figs.\,\ref{fig:fitmaps_ngc6860}, \ref{fig:fitmaps_mrk915} and \ref{fig:fitmaps_mcg0124012}, for more detailed contours, comparing the flux distribution of {\oiii} and CO(2-1).

\begin{figure*}
    \centering
    \includegraphics[width=1\linewidth]{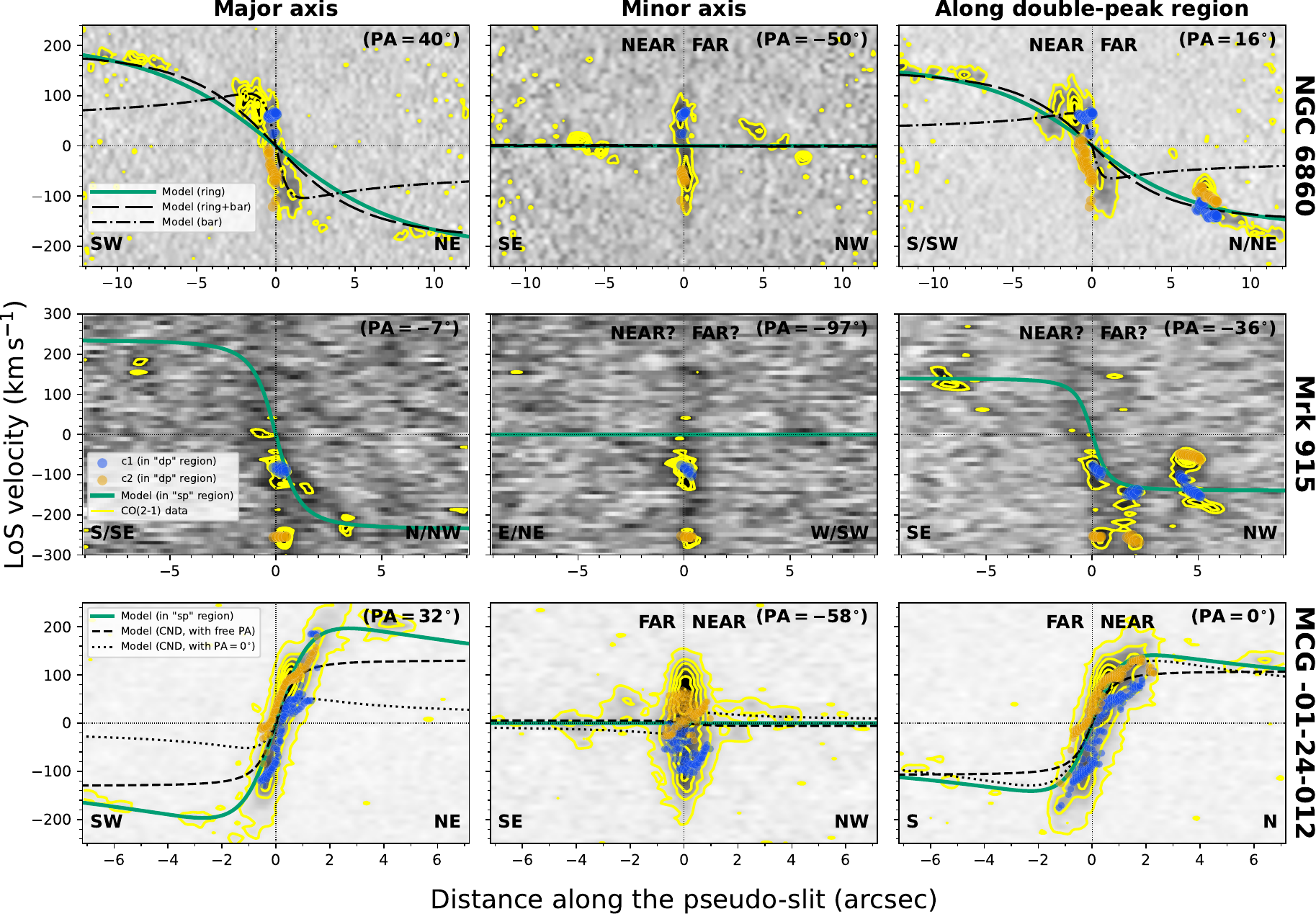}
    \caption{Position velocity (PV) maps for the CO(2-1) of each object for 1{\arcsec}-width pseudo-slits, along three different positions: kinematics major and minor axis (first two columns), and along the double-peak (dp) region (last column). We added the individual LoS velocities obtained for the Gaussian components \texttt{c1} (blue circles) and \texttt{c2} (orange circles) fitted to the spectral profiles. 
    The bluish-green lines correspond to the global rotating disk model fitted to the single-peak (sp) region (first three rows in Table\,\ref{tab:bertola2}). 
    For NGC\,6860, three models are shown, which were fitted using data from the ring (bluish-green line), from the bar (dot-dashed black line), and combining data from both ring and bar (long dashed black line).
    For MCG\,-01-24-012, we also added curves from two models fitted in the double-peak region, one letting the PA free (short dashed black line) and the other with PA\,=\,0{\degree} (dotted black line). 
    }
    \label{fig:pv}
\end{figure*}

\begin{figure*}
	\includegraphics[width=1\linewidth]{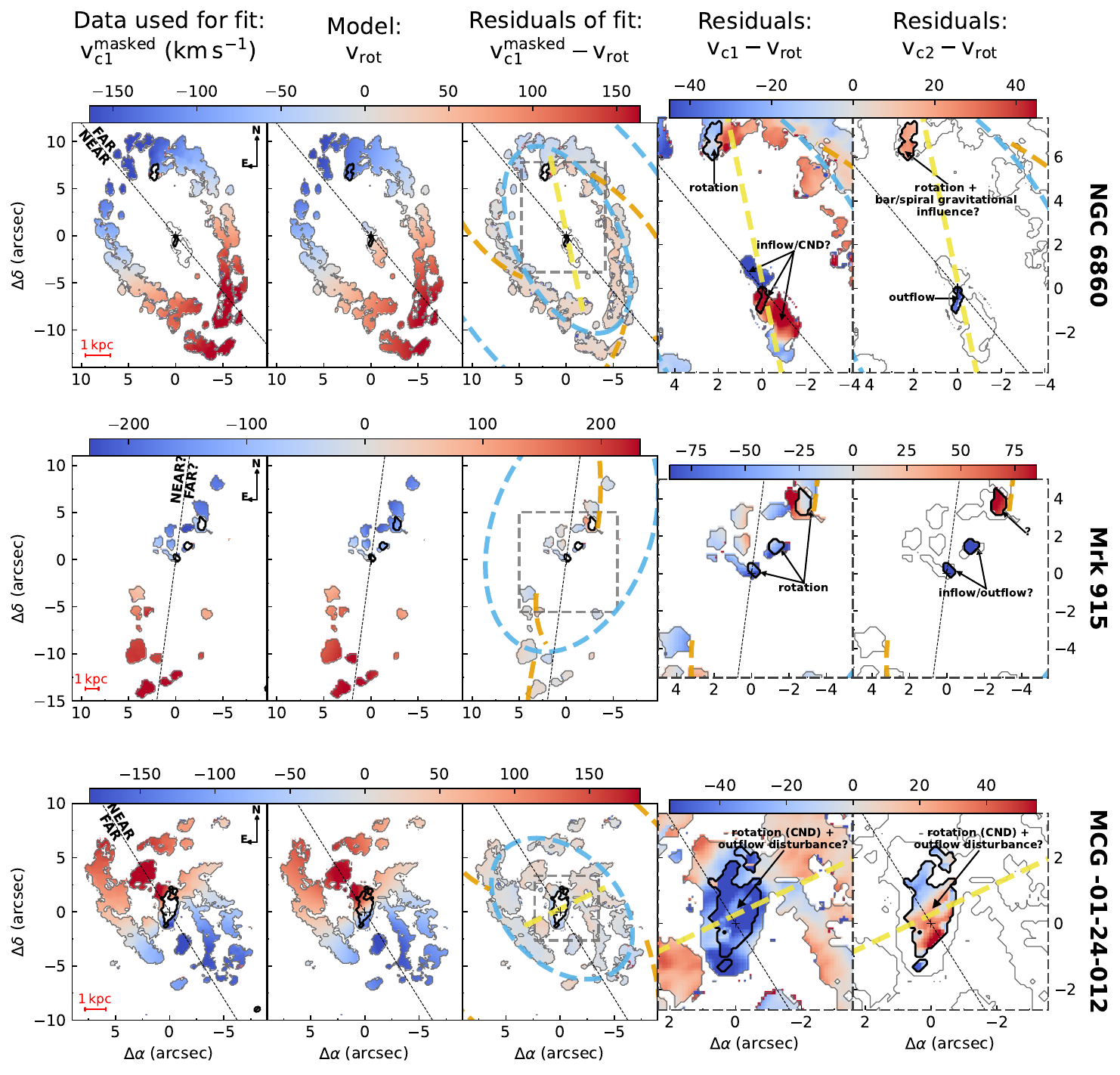}
    \caption{Comparison between the CO(2-1) data, obtained from the Gaussian decomposition of the profiles, and the rotating 2D disk model (first three rows in Table\,\ref{tab:bertola2}). The first three columns show the distributions of the LoS velocity of the \texttt{c1}-component ($\vel_{\tt{c1}}$), the best-fit rotation model, and the residuals from the fit. 
    The model was fitted to the single-peak region of the $\vel_{\tt{c1}}$ data (masked regions not shown). An additional mask was used for the inner regions of NGC\,6860, restricting the fit to regions dominated by rotation in the disk.
    The last two columns are a zoom-in showing residuals of each component (\texttt{c1} and \texttt{c2}), with the models extended to the double-peak region.
    The black contours encircle the masked regions with double-peak CO profiles, while the gray contours show the full data extent. 
    The dashed lines identify stellar structures: bars (yellow), spirals (orange), and rings (sky-blue).}    
    \label{fig:bertola}
\end{figure*}

\begin{figure*}
	\includegraphics[width=1\linewidth]{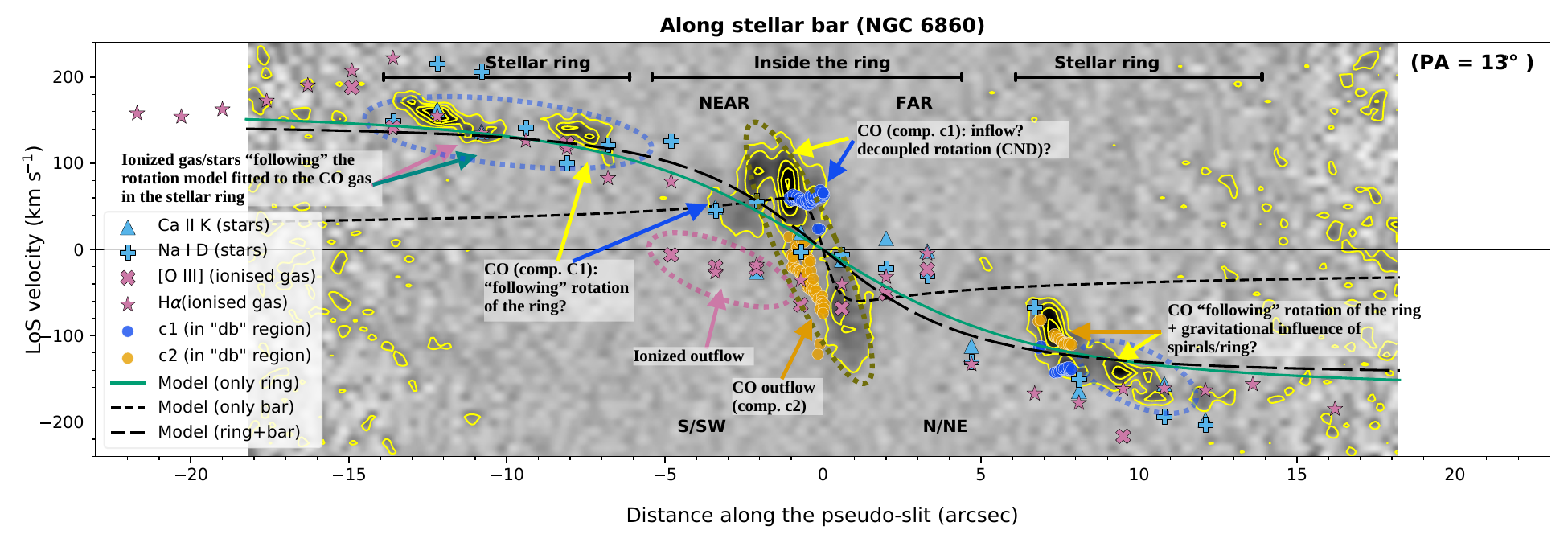}
    \caption{
        Position velocity (PV) map of NGC\,6860, similar to Fig.\,\ref{fig:pv}, but for a position angle along the bar and with a slit width of 1.5\arcsec, matching the observations from \citet{lipari+93}. We added their observed LoS velocities from: the stellar absorption lines of Ca\,II\,K (triangles) and Na\,I\,D (crosses) in reddish-purple color, and the ionized gas emission lines of {\oiii} (star symbols) and {\ha} (`x' markers) in sky-blue.}
    \label{fig:pv-ngc6860}
\end{figure*}

\section{Results: global rotation}\label{sec:results}

The bulk of the cold molecular gas is rotating in a disk, particularly in the outer regions where the profiles are narrower. 
The mean values of residuals in the single-peak CO region are $\sim$\,20\,--\,30\,{\kms}, and about $\sim$\,10 percent of the maximum LoS velocities in the regions with highest velocities in the disk model (see third column in Fig.\,\ref{fig:bertola}).
For comparison, using formulas from \citet{lenzs_ayres92}, we estimated a mean uncertainty in the fitted LoS velocities (Gaussian centers) of $\sim$\,2\,--\,3\,{\kms}, with maximum uncertainties of $\sim$\,10\,{\kms} in regions with low S/N. We believe the latter value better reflects the uncertainties, as it matches the channel widths of the cube. 
Therefore, the maximum residuals are a factor of $\sim$\,2\,--\,3 above the uncertainties. This might be a sign that, although our model for the global rotation reproduces well the kinematics in the single-peak regions, local forces also influence the molecular gas. Possible candidates might be local gravitational forces associated with stellar structures, since the CO is mostly observed close to stellar rings, bars, and spirals. Although not shown here, dust lanes might also play a role \citep{konig+14}. 

However, some regions present prominent distinct kinematics, reaching residual LoS velocity values of up to $\sim$\,100{\kms}, as shown by the PV diagrams in  Fig.\,\ref{fig:pv}. 
Along the kinematic minor axes, we observe non-zero velocity emission in all sources, which has been used as a signature of non-circular motions \citep[e.g.,][]{ramos-almeida+22}. These regions were better modeled using two Gaussian components (see the individual LoS velocity values (circles), reinforcing that the double-peak region is tracing disturbed CO kinematics.

In the following Sections\,\ref{sec:results_ngc6860}\,--\,\ref{sec:results_mcg0124012}, we analyze the three sources in our sample individually.

\section{Results: NGC\,6860}\label{sec:results_ngc6860}

\subsection{NGC\,6860: CO in the stellar ring}\label{sec:ngc6860_res_ring}

The CO emission in NGC\,6860 is observed along its inner stellar ring and bar. 
In particular, the CO gas kinematics in the ring is dominated by rotation (see Fig.\,\ref{fig:bertola}).

In addition, at $\sim$\,7.6{\arcsec} to the north of the NGC\,6860's nucleus (in the ring, at the intersection with the bar), there is a region with double-peak CO(2-1) profiles.
Both \texttt{c1} and \texttt{c2} components present low LoS velocity residuals relative to the disk model: less than $\sim$\,20\,{\kms} in absolute values (last two columns of Fig.\,\ref{fig:bertola}). Their residuals seem to follow the pattern of the gas in neighboring spaxels, with its kinematics probably being dominated by rotation. 
For the CO gas traced by these components, we assumed that: \texttt{c1} is only rotating, since it its lower relative residuals;  \texttt{c2} is rotating, but suffers a small local gravitational disturbance, probably introduced by the nearby spiral and/or the bar. 
From now on, we concentrate our analysis on regions closer to the nucleus.


\subsection{NGC\,6860: inflowing CO clouds in the stellar bar?}\label{sec:ngc6860_res_bar_inflow}

A stellar bar structure, with a position angle PA\,$\sim$\,12{\degree}, lies within the ring (yellow dashed line in Fig.\,\ref{fig:fitmaps_ngc6860}). Along this bar and inside a 4{\arcsec} radius ($\sim$\,1\,kpc), the CO LoS velocity distribution of the component \texttt{c1} shows high deviations relative to circular orbits (see Fig.\,\ref{fig:bertola}), with maximum residuals of $\sim$\,75\,{\kms}.  
Given the negative values on the far side, and mostly positive on the near side, these residuals suggest inflows along the bar.

The presence of molecular inflows along the bar in the inner kiloparsec was already proposed by \citet{lipari+93}. Their hypothesis was based on the possible presence of an ionized gas inflow, concluded after analyzing the velocity residuals of {\oiii} and {\ha} relative to a rotation disk model fitted to the data. However, they assumed the northwest as the near side of the galaxy, but here we could better constrain it to be in the southeast (see Sect.\,\ref{sec:dust}). 
Using our disk orientation, these ionized gas residuals should be interpreted as ionized gas outflow.

For a better understanding of the dynamics of NGC\,6880, we generated an additional PV diagram in Fig.\,\ref{fig:pv-ngc6860}, reproducing the slit configuration of \citet{lipari+93}. We added the LoS values\footnote{Approximate values, collected from Fig.\,7 of \citet{lipari+93}, using  PlotDigitizer: \url{https://plotdigitizer.com}} from the stellar Ca\,II\,K and  Na\,I\,D absorption lines and ionized gas emission lines {\oiii} and {\ha}, together with our CO data and its kinematic models. Overall, the stellar and the ionized gas seem to follow the rotation model of CO in the ring. However, in the lower left quadrant, under 5{\arcsec} of the nucleus (reddish-purple dotted line), the ionized gas shows velocity residuals. These are the residual originally detected by \citet{lipari+93}, which we have reinterpreted as coming from an ionized outflow (as discussed in the previous paragraph).

Assuming a constant gas inflow rate along the bar and an inclination of $i$\,=\,57.7{\degree} (co-planar with the galaxy disk), we obtained a cold molecular mass inflow rates of $\dot{M}_{\rm{mol,in}}$\,$\sim$\,0.8\,--\,6\,{\msunyr} (see Table\,\ref{tab:inflow}).
For an inclination of $i$\,=\,70{\degree}, the $\dot{M}_{\rm{mol,in}}$ value would decrease by a factor of $\sim$\,1.7, while for $i$\,=\,40{\degree}, it would increase by a factor of $\sim$\,1.9.
The total inflowing molecular gas mass involved in the process is $M_{\rm{mol,in}}$\,$\sim$\,(1\,--\,9)\,${\times}$\,10$^{7}${\msun}, corresponding to volumetric number density of molecular clouds of $n_{\rm{mol}}$\,$\sim$\,3\,--\,30\,cm$^{-3}$.
See Appendix\,\ref{ap:inflow}, for details on the calculation.

This range of $n_{\rm{mol}}$ values is compatible with the results obtained for diffuse clouds by \citet{luo+23}. 
The presence of a more diffuse gas in the bar could be a consequence of the bar-induced turbulence and even help to suppress star formation inside the ring \citep[e.g.,][]{khoperskov+18,scaloni+24}. 
And indeed, the {\oiii}/(\ha+\nii) excitation map from  \citet{lipari+93} indicates AGN-like ionization in this region, without signs of star formation. 
Another proposed explanation is that the bar creates an inflow of gas, caused by a strong torque that drives the gas directly to the galactic center, where in some cases star formation is enhanced \citep[e.g.,][]{spinoso+17}. 
However, we note that there is no evidence of nuclear star formation in NGC\,6860 \citep{lipari+93}. 

Higher turbulence in the CO can also be induced by shear in the inflowing gas along dust lanes, as observed in the minor merger systems NGC\,4194  \citep{konig+13} and NGC\,1614 \citep{konig+14}. Similarly to NGC\,6860, ongoing star formation is not observed in these turbulent regions. However, part of the inflowing material in these two systems -- probably associated with the recent galaxy interaction -- ends up fueling ongoing star formation in other regions nearby, which does not seem to be the case for NGC\,6860. Besides that, as shown in Fig.\,\ref{fig:companions} and pointed out in Sect.\,\ref{sec:sample_ngc6860}, there is no clear sign of a recent merger in NGC\,6860, although deeper broadband imaging observations may reveal a different picture. 
We cannot discard that part of the inflowing material will not be used to form new stars in the future, or that there is deeply embedded star formation happening in the central region.
In Sect.\,\ref{sec:disc_inflow}, we show that only a fraction of the inflowing gas ends up feeding the AGN, and discuss whether we should expect to observe a reservoir of CO accumulating in the nucleus. 

\subsubsection{NGC\,6860: alternative scenario for the CO in the bar: decoupled disk rotation?}\label{sec:ngc6860_res_bar_CND}

There is an alternative scenario to explain the CO kinematics along the bar, within $\sim$\,1\,kpc from the nucleus: a decoupled rotation from the gas in the ring.  This hypothesis was raised because the steep LoS velocities in the PV diagrams (Figs.\,\ref{fig:pv} and \ref{fig:pv-ngc6860}) seem to be characteristic of a rotation (instead of inflows). 
Different scenarios have been proposed to explain a decoupled kinematics between the gas in the inner and outer regions, including infalling of new gas material from cosmological filaments and mergers with gas-rich dwarfs \citep[e.g.,][]{thakar+96}. 
In these cases, the decoupled kinematics arise from a misalignment between the disk angular momentum of the gas in the galaxy disk and the arriving material.

To test this, we modeled the gas kinematics in this region with an independent rotating disk model \citep{bertola+91}, fixing the systemic velocity but leaving the remaining parameters free. And, indeed, a rotating gas model can fit reasonably well the LoS velocities in the bar, as shown by the corresponding parameters maps in Fig.\,\ref{fig:bertola-ngc860_onlybar} (in Appendix\,\ref{ap:additional_fig}) and the dashed-dotted black lines in the PV diagrams (Figs.\,\ref{fig:pv} and \ref{fig:pv-ngc6860}).
The best-fit parameters (see in Table\,\ref{tab:bertola2}) show a similar position angle to that of the ring (PA\,$\sim$\,40{\degree}), but with the plane of rotation having a higher inclination (i\,$\sim$\,69\degree).

The elongated CO emission distribution in the bar supports an internal inclined disk, although one might argue that this scenario would also require that the major axis be more aligned with the bar (PA\,$\sim$\,12\degree). 
Another argument against this alternative scenario is that the velocities of the stars and the ionized gas -- in the bar -- agree with the rotation model of CO in the ring (see Fig.\,\ref{fig:pv-ngc6860}). However, stars, ionized, and cold molecular gas do not necessarily need to have the same rotation pattern. The ionized and molecular gas can differ, for example, in the amplitude of the rotation velocity and the dispersion of the velocity \citep[e.g.,][]{levy+18}. 
In the more pronounced cases, counter-rotation between gas and star is present \citep[e.g.,][]{garcia-burillo+00,garcia-burillo+03}.

Since we cannot fully discard any of the above scenarios, we considered both explanations for gas in the bar traced by the component \texttt{c1}: inflows or decoupled rotation. 
Finally, we note that the latter hypothesis could be interpreted as a circumnuclear disk (CND), similar to the one in MCG\,-01-24-012 (see Sect.\,\ref{sec:mcg0124012_res_doublepeak}).


\subsection{NGC\,6860: a molecular outflow encasing the ionized one}\label{sec:ngc6860_res_outflow}

There is another CO double-peak region, close to the nucleus and within a $\sim$\,1{\arcsec} radius (see Fig.\,\ref{fig:fitmaps_ngc6860}). This region is located at the intersection between the {\oiii} bipolar cone and the CO gas distribution. The kinematics of the \texttt{c1} component in this region is similar to the gas in the single-peak region around it, indicating that it is tracing inflowing gas or the CND motion, depending on the interpretation for the CO in the bar (see Sects.\,\ref{sec:ngc6860_res_bar_inflow} and \ref{sec:ngc6860_res_bar_CND}). Independent of the interpretation for \texttt{c1}, the emission from the \texttt{c2} component is likely tracing outflowing molecular clouds: 
it shows negative residuals relative to the global disk rotation model, which are located at the near side of the disk (see Fig.\,\ref{fig:bertola}, last column).

These outflowing molecular clouds are located at the border of the {\oiii} bipolar emission. This elongated ionized gas emission also has evidence of outflows from optical long-slit observations along the east-west direction \citep[PA\,=\,85\degree]{bennert+06}: the ionized gas has negative LoS velocities at the near side and positive at the far side of the disk. In addition, the line ratios in the nuclear region are  AGN-like \citep{bennert+06,lipari+93}. 
Therefore, in the nuclear region, the ionized gas outflow seems to be partly surrounded by a molecular one,  traced by the \texttt{c2} component in our CO(2-1) data, as represented in the first column of Fig.\,\ref{fig:models}. 

This suggests that in NGC\,6860 the outflowing molecular clouds can only survive at the borders of the ionized cones, being destroyed closer to the ionization axis.  This scenario is similar to what we observed in the Seyfert galaxy NGC\,3281 \citep{dallagnol+23}. Other objects presenting outflowing molecular clouds with a similar hour-glass morphology --  although sometimes surrounding X-ray cavities or radio lobes -- include the Milky Way \citep[e.g.,][]{veena+23}, starburst galaxies \citep[e.g.,][]{bolatto+13b}, radio jets/bubbles \citep[e.g.,][]{morganti+23,zanchettin+23}, galaxy clusters \citep[e.g.,][]{russel+17}, and even newborn stars \citep[e.g.,][]{delabrosse+24,nagar+97,zhang+19}. The existence of such scenarios in quite different astronomical environments suggests that these ``onion-like'' outflows might be common. 

The above hypothesis relies on the assumption that outflows observed in both phases were launched at the same time. Another possible scenario would be that the CO clumps were ejected before, with the ionized gas being launched later and filling the gap left behind by the other phase. In this scenario, one might expect that the cold molecular outflow extent ($\sim$\,1\arcsec) should be larger than the ionized one ($\sim$1.3\,\arcsec), which does not seem to be the case. However, this is based on the assumption that both phases have the same outflow velocities and inclinations, which is not necessarily true. 
The opposite case, where the ionized gas had been ejected earlier, can also be considered. 
Besides that, we cannot discard that we might be missing part of the total molecular outflow content, which might be present in denser (e.g., as traced by HCN molecular lines), hotter phases (e.g.,  H$_2$ near-infrared lines), and/or other CO isotopes (e.g., $^{13}$CO lines).
Therefore, even though we favor a scenario where both the ionized and the cold molecular gas are being ejected together, with the latter only surviving at the border of the ionized cone, we cannot fully discard that other scenarios might better explain the observations.


Following Appendix\,\ref{ap:outflow}, the total molecular mass of the outflowing clouds in NGC\,6860 is $M_{\rm{mol,out}}$\,$\sim$\,$0.6$\,--\,$5$\,$\times$\,$10^6$\,{\msun} (see Table\,\ref{tab:outflow}). This corresponds to $\sim$\,0.7 percent of the total molecular mass in the galaxy observed within a radius of $\sim$\,13{\arcsec}. 
Assuming that outflow is co-planar to the galaxy disk ($i$\,=\,57.7{\degree}), the de-projected maximum velocity and outflow extent are $V_{\rm{out}}$\,$\sim$\,140\,{\kms} and $R_{\rm{out}}$\,$\sim$\,560\,pc, respectively, with the molecular mass outflow rate being $\dot{M}_{\rm{mol,out}}$\,$\sim$\,0.1\,--\,1\,{\msunyr}. 
For a different inclination of $i$\,=\,70{\degree} ($i$\,=\,40{\degree}), the $\dot{M}_{\rm{mol,out}}$ value would decrease (increase) by a factor of  $\sim$\,1.7 ($\sim$\,1.9).

\begin{table}
    \centering
    \caption{
    Inflow properties} 
\begin{tabular}{ccccccc}
\hline\hline
 Name & $M_{\rm{mol,in}}$ & V$_{\rm{in}}$ &  $\dot{M}_{\rm{mol,in}}$ \\
      & $10^7$\,{\msun}   & \kms          &  {\msunyr}               \\
 (1) & (2) & (3) & (4) \\
\hline
 NGC\,6860$^a$ & 
 1\,--\,9
 & 90  & $0.8$\,--\,$6$     \\
 Mrk\,915$^b$  & 0.09\,--\,0.8 & 130 & $0.09$\,--\,$0.8$  \\
\hline
\end{tabular}
\tablefoot{
    (1) Name;
    (2) Inflowing molecular mass (sum between both regions marked in Fig.\,\ref{fig:inflow}) and velocity (mean from (3) both regions); 
    (4) Mass inflow rate.  
    See Sects.\,\ref{sec:ngc6860_res_bar_inflow} and \ref{sec:inflow_mrk915_leading}, and Appendix\,\ref{ap:inflow}, for additional information.
    \\
$^{a}$: Assuming that the CO gas along the NGC\,6860's bar is inflowing. \par
$^{b}$: Assuming that the spiral arms of Mrk\,915 are leading  (see Appendix\,\ref{ap:mrk915-orientation}). \par
}
\label{tab:inflow}
\end{table}

\begin{table*}
    \centering
    \caption{
    Outflow properties
    }
\begin{tabular}{ccccccc}
\hline\hline
Name &  $S_{\nu} \Delta \vel_{\rm{CO(2-1),out}}$ & $M_{\rm{mol,out}}$ & $f_{\rm{out}}$ & $V_{\rm{mol,out}}$    & $R_{\rm{mol,out}}$ & $\dot{M}_{\rm{mol,out}}$ \\
     & {\mjykms}                                 & $10^7$\,{\msun}    & \%             & $\mathrm{km\,s^{-1}}$ & $\mathrm{kpc}$     & {\msunyr} \\
(1) & (2) & (3) & (4) & (5) & (6) & (7) \\
\hline
NGC\,6860       &  $340\,{\pm}\,6$$^{b}$   & $0.06$\,--\,$0.5$ & 0.7 & 140  & 0.56 & $0.1$\,--\,$1$ \\
Mrk\,915$^{a}$  &  $184\,{\pm}\,4$$^{c}$   & $0.08$\,--\,$0.7$ & 4   & 300 & 2.8  & $0.09$\,--\,$0.7$ \\
MCG\,-01-24-012 &  $9220\,{\pm}\,10$$^{d}$ & $3$\,--\,$20$     & 30  & 40  & 3  & $0.4$\,--\,$3$    \\
\hline
\end{tabular}
\tablefoot{ 
    (1) Galaxy name; 
    (2) Disturbed/outflowing CO(2-1) flux; 
    (3) Disturbed/outflowing cold molecular gas mass; 
    (4) Fraction of molecular mass that is perturbed: $f_{\rm{out}}$\,=\,$M_{\rm{mol,out}}/M_{\rm{mol,tot}}$; 
    (5) De-projected outflow velocity; 
    (6) De-projected outflow radius; 
    (7) Mass outflow rate. 
    The inclinations used were 57.7{\degree}, 66.5{\degree} and 74{\degree} for NGC\,6860, Mrk\,915, and MCG\,-01-24-012, respectively.
    See Sects.\,\ref{sec:ngc6860_res_outflow}, \ref{sec:outflow_mrk915_trailing} and \ref{sec:mcg0124012_res_doublepeak}, and Appendix\,\ref{ap:outflow}, for additional information.\\ 
$^{a}$: Assuming that the spiral arms of Mrk\,915 are trailing (see Appendix\,\ref{ap:mrk915-orientation}). \par
$^{b}$: Flux integrated from the \texttt{c2}-component, inside a 0.6{\arcsec} radius. \par
$^{c}$: Flux integrated from the \texttt{c2}-component, inside a 2.3{\arcsec} radius.  \par
$^{d}$: Integrated from both components in the double-peak region. \par
}
\label{tab:outflow}
\end{table*}

\section{Results: Mrk\,915}\label{sec:results_mrk915}

\subsection{Mrk\,915: CO along the spiral arms}\label{sec:mrk915_res_spirals}

The CO emission in Mrk\,915 is distributed along its two inner spiral arms, extending up to a $\sim$\,15{\arcsec} radius. 
Along the northwest spiral arm, we detected double-peak CO profiles in Mrk\,915 in three regions spread approximately along a projected line (at PA\,$\sim$\,-36{\degree}), starting at the nucleus and extending up to $\sim$\,5{\arcsec} to the northwest (see Fig\,\ref{fig:fitmaps_mrk915}).
Both \texttt{c1} and \texttt{c2} components in the two inner CO clumps show negative LoS velocities relative to the disk rotation model (middle column in Fig.\,\ref{fig:fitmaps_mrk915}). 
Since the absolute LoS velocity residuals values of the \texttt{c1} component are lower than those of \texttt{c2} (Fig.\,\ref{fig:bertola}), we assumed that the \texttt{c1} component in the double-peak region is tracing gas rotating in the disk.

\subsection{Mrk\,915: CO in the double-peak region}\label{sec:outflow_mrk915_db}

The interpretation of the kinematics of the component \texttt{c2} depends on the orientation of the disk, which we already noted is ambiguous (see Appendix\,\ref{ap:mrk915-orientation}). 
Schematic models considering the two possible orientations are shown in Fig.\,\ref{fig:models}, including the resulting spiral arms motion: trailing (second column) and leading (third).
Independent of the orientation, the LoS velocity residuals are complex, showing negative values closer to the nucleus (covering the two nearest CO clumps under a $\sim$\,2.3{\arcsec} radius) and positive at the farthest clump (at $\sim$\,5{\arcsec} radius). 

We chose to ignore the farthest CO cloud in our analysis.
The decision was a consequence of the lack of a clear scenario that could properly explain the kinematics of the three \texttt{c2}-clumps in the double-peak region.
This could be a consequence of the low S/N spectra of Mrk\,915. Deeper observations might allow better modeling of disk rotation, yielding different LoS velocity residual values.

\subsubsection{Mrk\,915: molecular inflow (leading arms scenario)?}\label{sec:inflow_mrk915_leading}

If we assume that the near side is to the east (from the dust attenuation method, leading to a leading spiral arms scenario), the origin of the negative residuals of $\vel_{\tt{c2}}$ (from the two inner CO clumps) could be interpreted as coming from inflowing molecular clouds. 

Following Appendix\,\ref{ap:inflow}, we obtained a molecular mass inflow rate along the one of the spirals of $\dot{M}_{\rm{mol,in}}$\,$\sim$\,0.09\,--\,0.8\,{\msunyr}, involving a total molecular mass of $M_{\rm{mol,in}}$\,$\sim$\,(0.9\,--\,8)\,${\times}$\,10$^{6}${\msun}. This corresponds to volumetric number density of $n_{\rm{mol}}$\,$\sim$\,0.4\,--\,3\,cm$^{-3}$, which is an order of magnitude lower than the value obtained for NGC\,6860. 
We assumed an inflow inclination of $i$\,=\,66.5{\degree} (co-planar with the disk), for these calculations.
For an inclination of $i$\,=\,80{\degree} ($i$\,=\,50{\degree}), the $\dot{M}_{\rm{mol,in}}$ value would decrease (increase) by a factor of  $\sim$\,2.5 ($\sim$\,1.9). The same dependence on the inclination applies for mass outflow rate value (see this scenario in Sect.\,\ref{sec:outflow_mrk915_trailing})

Given that Mrk\,915 is part of a triple system (see Sect.\,\ref{sec:sample_mrk915} and Fig.\,\ref{fig:companions}), the gravitational influence of the two bright companions might have disturbed the gas in the host galaxy disk. If this gravitational influence leads to a significant loss of angular momentum in the gas, we may be currently observing the resulting CO inflow motion. 
We also note that, since both components are observed along the same LoS in the double-peak region, the ``non-rotating'' \texttt{c2} might be tracing out-of-plane gas, which could be a consequence of material flowing from tidal streams of gas from these nearby companions.

\subsubsection{Mrk\,915: molecular outflow (trailing arms scenario)?}\label{sec:outflow_mrk915_trailing}

On the other hand, if the near side is to the west (agreeing with trailing arms), the origin of the \texttt{c2} emission could be outflowing clouds. The large negative values of $\vel_{\tt{c2}}$ seem to favor this scenario, with mean projected residuals values of $\sim$\,$-220$\,{\kms} in the nucleus. 
There is also a small elongation in the 3.6\,cm radio, although along another direction (PA\,$\sim$\,220\degree), that could be associated with a radio jet (see Fig.\,\ref{fig:models}). However, contrary to our other two sources, the {\oiii} emission is not bipolar, which would help identify a possible ionization axis. 

Following this scenario (see Appendix\,\ref{ap:outflow}, for details), the total cold molecular mass of the outflowing clouds is $M_{\rm{mol,out}}$\,$\sim$\,$0.8$\,--\,$7$\,$\times$\,$10^6$\,{\msun}, corresponding to 4\,percent of the total mass of the gas in this phase. If we assume an outflow co-planar to the disk ($i$\,=\,66.5{\degree}), we obtain a molecular mass outflow rate of $\dot{M}_{\rm{mol,out}}$\,$\sim$\,0.09\,--\,0.7\,{\msunyr}, for an deprojected outflow velocity of $V_{\rm{out}}$\,$\sim$\,300\,{\kms} and radius of  $R_{\rm{out}}$\,$\sim$\,2.8\,kpc.
If we vary the inclination by $\sim$\,15{\degree}, the $\dot{M}_{\rm{mol,out}}$ value would change by a factor of $\sim$\,2, analogous to the effect on $\dot{M}_{\rm{mol,in}}$ (shown in Sect.\,\ref{sec:inflow_mrk915_leading}). 

In the schematic models shown in Fig.\,\ref{fig:models}, we added two {\oiii} cones aligned with the {\oiii} emission. They were drawn assuming that the asymmetric {\oiii} elongation and the 3.6\,cm emission radio are possibly tracing two distinct ionization axes, and considering that they could have partly destroyed the molecular gas. Maybe one of the {\oiii} elongations could be a relic of a previous AGN orientation, possibly associated with a galaxy interaction. 
In this case, the {\oiii} flux distribution would be a consequence of two distinct accretion events. Supporting this, we found signs of a possible tidal trail between Mrk\,915 and one of its two nearby galaxy companions, as noted in Sect.\,\ref{sec:sample_mrk915} (see Fig.\,\ref{fig:companions}). 
Nonetheless, we emphasize that this is just a speculative scenario, and we don't have strong evidence in favor of it.
Resolved integral field observations of the ionized gas should help to clarify the geometry and kinematics of the {\oiii} emitting gas.

\begin{figure}
	\includegraphics[width=1\linewidth]{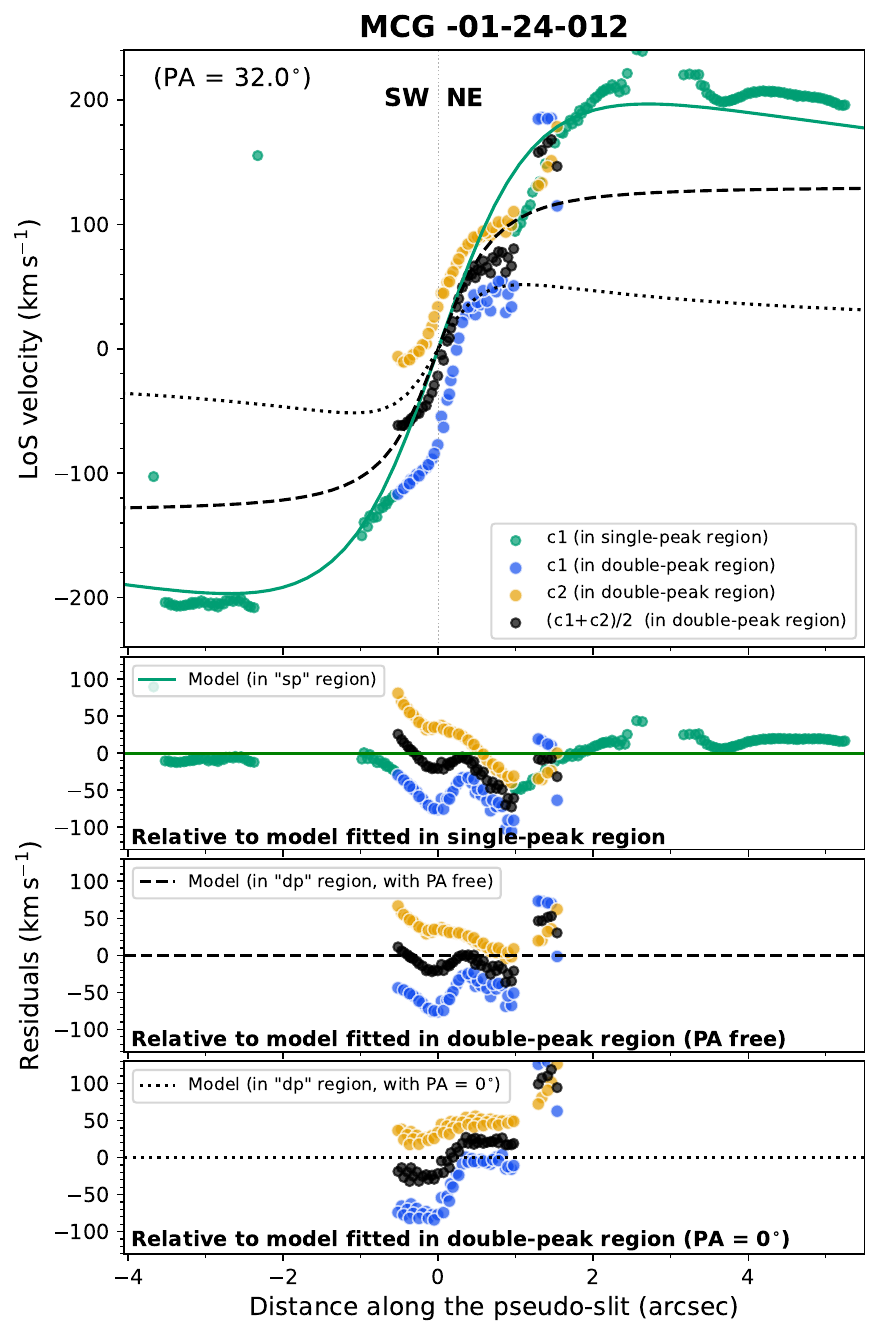}
    \caption{The top row shows the distribution of line of sight velocity values versus the distance along the global kinematic major axis (PA\,=\,32.0\degree, inside a 0.27{\arcsec} slit width) of MCG\,-01-24-012. Values in the single-peak region are shown as blueish-green circles (component \texttt{c1}), while values in the double-peak region are shown as blue (for \texttt{c1}) and orange circles (for \texttt{c2}). The black circles correspond to the average between the LoS velocities of both components ($\vel_{\rm{rot,CND}}$). 
    The global disk model, obtained in the single-peak region (see Fig.\,\ref{fig:bertola}), is shown as a blueish-green line, with the residual shown in the second row. We show two models fitted in the CND region (see maps in Fig.\,\ref{fig:bertola-mcg-inner-vel}, and schematic models in Fig.\,\ref{fig:models_mcg}): one with the a free PA parameter (black dashed line, with the residuals in the third row); and another with PA\,=\,0{\degree} fixed (black dotted line, with the residuals in the forth row).}
    \label{fig:bertola-mcg-inner-radial}
\end{figure}

\begin{figure}
	\includegraphics[width=1\linewidth]{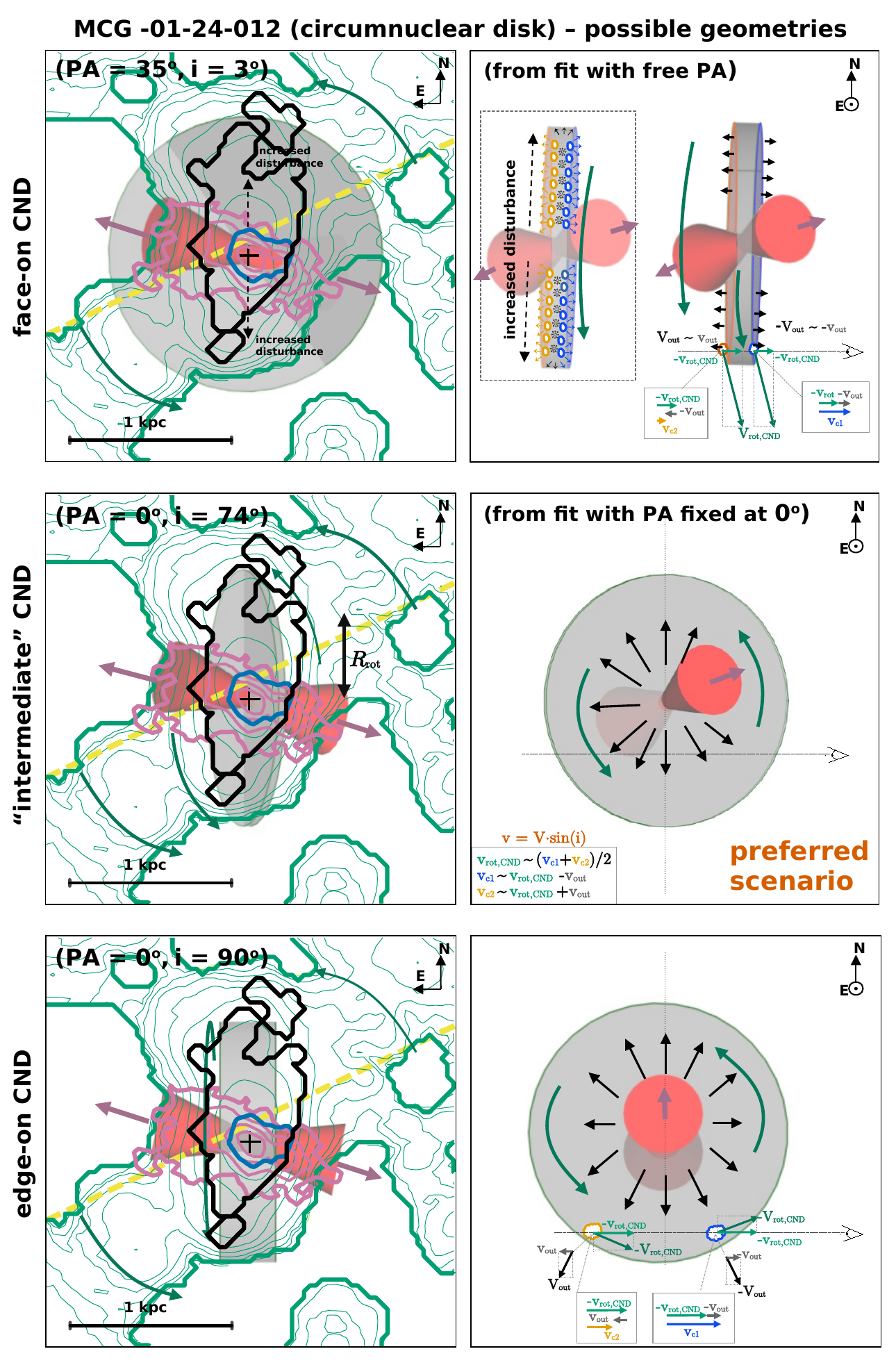}
    \caption{
    Possible geometries for the CND in MCG\,-01-24-012. The legend from Fig.\,\ref{fig:models} also applies here. 
    The left and right columns show the observation and lateral views, respectively. On the right, we added arrows to represent the velocity vectors of the components \texttt{c1} (blue) and \texttt{c2} (orange), decomposed in the rotating (bluish-green) and outflowing (black) parts, with diagrams. As in Fig.\,\ref{fig:models}, the orientations of the ionized cones were chosen to try to reproduce the {\oiii} emission (reddish-purple contours). 
    In the first row, the face-on CND is the result of fitting a rotating disk model with free PA. In this scenario, the CO double peaks would result from expanding clouds perpendicular to the disk. 
    In the last row, we show another way of reproducing the double peaks by assuming an edge-on disk geometry, with the outflow occurring radially along the disk. Our preferred scenario lies between these two geometries, as shown in the middle row. This model is the result of fixing PA at 0{\degree} to reproduce the CO flux distribution, which yields an inclination of 74{\degree}. 
    }
    \label{fig:models_mcg}
\end{figure}

\section{Results: MCG\,-01-24-012}\label{sec:results_mcg0124012}

\subsection{MCG\,-01-24-012: CO in the stellar ring and bar}\label{sec:mcg0124012_ring_bar}

The observed cold molecular gas -- as traced by CO(2-1) -- is observed along the ring and inside its radius $\sim$\,7.5{\arcsec} (see Fig.\,\ref{fig:fitmaps_mcg0124012} and \ref{fig:models}). 
Along the ring, the CO kinematics is dominated by rotation, as shown in the residuals relative to the disk model (Fig.\,\ref{fig:bertola}).

Unlike NGC\,6860, it is not clear whether the bar is affecting the molecular gas kinematics in MGC\,-01-24-012. 
The presence of a bar tends to ``clean'' its vicinity inside the co-rotational (ring) radius, concentrating the gas close to the bar, the ring, and the nucleus. This might be the case of NGC\,6860, which has a more prominent bar, but not of MCG\,-01-24-012, where the CO emission is observed in regions -- inside the ring -- that are not co-spatial with the bar. For example, the brightest CO structure inside the ring (elongated emission along PA\,$\sim$\,0{\degree}) is tilted relative to the bar (PA\,$\sim$\,115{\degree}), being observed extending by $\sim$\,1.5{\arcsec} to the north and the south of the nucleus. 
Maybe we are witnessing the stellar bar being formed, with its local effects on the gas not well established yet.

\subsection{MCG\,-01-24-012: CO in a circumnuclear disk being disturbed by an ionized outflow?}\label{sec:mcg0124012_res_doublepeak}

In the double-peak region, both components show LoS velocity distribution with patterns that seem to follow the overall rotation of the disk (middle maps in Fig.\,\ref{fig:fitmaps_mcg0124012}). There are, however, significant residuals in LoS velocities relative to the disk model, with absolute residual values of up to $\sim$\,100\,{\kms}. 

To help visualize that, we present, in Fig.\,\ref{fig:bertola-mcg-inner-radial} (top panel), the LoS velocities as a function of distance (along the major axis of the outer galaxy disk) for each of the CO(2-1)'s components. The LoS radial velocity curve of each component, in the double-peak region, seems typical of gas rotating in a disk but with an additional negative (in $\vel_{\tt{c1}}$) and positive (in $\vel_{\tt{c2}}$) velocity shifts. Their radial profiles seem to differ from those of the gas in the single-peak region at the outer parts. This suggests the presence of a circumnuclear disk (CND) with a different orientation/geometry. 

We can approximately isolate the contribution from the rotation in the double-peak region by averaging the LoS velocities of both components: $\vel_{\rm{rot,CND}}$\,$\sim$\,($\vel_{\tt{c1}}$+$\vel_{\tt{c2}}$)\,/\,2 (black circles in Fig.\,\ref{fig:bertola-mcg-inner-radial}). Assuming that $\vel_{\rm{rot,CND}}$ traces the velocity of the rotation kinematics of the circumnuclear disk, we modeled it using a disk model from \citet{bertola+91}. More specifically, we fitted $\vel_{\rm{rot,CND}}$ in the double-peak region. The kinematic center was fixed again in the millimeter continuum peak, and the systemic velocity was fixed at the value obtained in the fit done for the single-peak region. The best-fit parameters correspond to a face-on disk geometry  (i\,$\sim$\,3\degree). The fourth row of Table\,\ref{tab:bertola2} displays the parameter values, with Fig.\,\ref{fig:bertola-mcg-inner-vel} (first row) showing the corresponding maps.

We observe that this face-on rotating disk model better reproduces the $\vel_{\rm{rot,CND}}$ curve compared with the model fitted in the single-peak regions, as highlighted by their model residuals Fig.\,\ref{fig:bertola-mcg-inner-radial} (second and third panels).
We sketched a possible scenario for such disk geometry in the first row of Fig.\,\ref{fig:models_mcg}: gas previously rotating is disturbed by an ionized outflow, with the energy propagating along this disk (in the north-south direction) and making the CO gas in the CND expand in all directions.
The double-peak profile would arise from the simultaneous observation of clumps in the front (with a negative velocity gain in $\vel_{\tt{c1}}$) and the back side of the disk (with positive gain in $\vel_{\tt{c2}}$) (see the lateral view on upper the right panel in Fig.\,\ref{fig:models_mcg}).

Nonetheless, it is unclear how a face-on disk could produce the observed CO central flux distribution, which extends along the north-south direction. 
Additionally, the resulting maximum velocity of the rotation model is unreasonably high: V$_{\rm{max}}$\,$\sim$\,2300\,{\kms} (see Table\,\ref{tab:bertola2}). In the last row of Fig.\,\ref{fig:models_mcg} we show another geometry for the CND that could explain the CO flux distribution and the double-peak profiles: an edge-on disk, containing originally rotating molecular clouds, that are outflowing radially -- along the CND -- due to an almost perpendicular ionized outflow. 
However, forcing an edge-on disk geometry for the disk ($i$\,$\sim$\,90{\degree} and PA\,$\sim$\,0{\degree}) resulted in a poor fit.

Therefore, we re-fitted the data, but fixing only PA at 0{\degree}, to better reproduce the CO flux distribution in the CND region. The fit returned an intermediate scenario with an inclination of $i$\,=\,74\,$\pm$\,2{\degree}. 
This time, the maximum velocity was lower: V$_{\rm{max}}$\,=\,270\,$\pm$\,40\,{\kms}. 
Since we fixed PA, the residuals are worse than the model with free PA.
The corresponding maps are shown in Fig.\,\ref{fig:bertola-mcg-inner-vel} (bottom row), while Fig.\,\ref{fig:bertola-mcg-inner-radial} shows its velocity radial plot (dotted black line) with the residual of each model being displayed in the last three rows.  
Since the residuals are more or less aligned with the stellar bar, we question if this stellar structure might also influence the gas in the CND, but we could not answer this with the current data.
We point out that the V$_{\rm{max}}$ value is of the order of the maximum velocity found for the circumnuclear disk in NGC\,1068, another barred Seyfert galaxy with a previously identified CND \citep{garcia-burillo+19}, supporting a similar scenario in MCG\,-01-012-24.

We conclude that this latter geometry -- with PA\,$\sim$\,0{\degree} and $i$\,$\sim$\,74{\degree} (middle row in Fig.\,\ref{fig:models_mcg}) -- is the best scenario for our observational constraints, as it both produces a rotation curve typical of CNDs and can also explain the flux distribution of the CO emission which is elongated along the north-south direction. 
Finally, although the CND could be fueling the AGN, we did not find clear evidence of it with current observations, except for the fact that the CND is approximately perpendicular to the ionized gas outflow, as expected for accreting structures around SMBH.

Based on the CND model fitted in the double-peak region, we proposed that the CO-emitting gas in the double-peak region was disturbed and partly pushed to the side by the passage of the {\oiii} ionized outflow, while keeping its bulk rotation motion. This interpretation is consistent with optical, near-infrared, and millimeter observations that report similar line profiles -- associated with higher velocity dispersions -- in regions outside/perpendicular to the AGN ionization cones and/or radio jets.
This has been observed both in the ionized gas \citep[e.g.,][]{venturi+21,riffelRA+21,finlez+18,lena+15,riffelRA+14,couto+13,finlez+18}, as well as in both warm \citep[e.g.,][]{riffelRA+15,diniz+15} and cold molecular gas \citep[e.g.,][]{ramos-almeida+22,shimizu+19,finlez+18}.  
Arguing in favor of this scenario, we note that MCG\,-01-24-012 also presents an elongated radio emission (although barely resolved), almost aligned with the {\oiii} emission. However, this is only weak evidence of a radio jet. Higher-resolution radio observations are needed to confirm it.

In the CND scenario, the total disturbed molecular mass obtained from CO is $M_{\rm{mol}}$\,$\sim$\,$3$\,--\,$20$\,$\times$\,$10^7$\,{\msun}, which corresponds to $\sim$\,30 percent of the total cold molecular mass. 
Assuming an outflow close to the CND plane ($i$\,=\,74{\degree}), the cold molecular mass outflow rate is $\dot{M}_{\rm{mol,out}}$\,$\sim$\,0.4\,--\,3\,{\msunyr}. 
For an different inclination of $i$\,=\,85{\degree} ($i$\,=\,60{\degree}), the $\dot{M}_{\rm{mol,in}}$ value would decrease (increase) by a factor of  $\sim$\,3.3 ($\sim$\,2.0). We used a de-projected outflow velocity and radius of $V_{\rm{out}}$\,$\sim$\,40\,{\kms} and $R_{\rm{out}}$\,$\sim$\,3\,kpc, respectively.
See Appendix\,\ref{ap:outflow}, for details on the calculation.

\section{Discussion: inflows and outflows}
\label{sec:Discussion}

NGC\,6860, Mrk\,915, and NGC\,6860 are part of a larger AGN sample of 13 nearby sources observed with ALMA, as described in \citetalias[][]{dallagnol+25_paperI}. Analogously to our three objects, in the other ten objects, the bulk of the CO gas is rotating in a disk, with all of them displaying local non-circular motions within a few kiloparsecs \citep{ramakrishnan+19,slater+19,finlez+18,salvestrini+20,dallagnol+23}, which could be expected since they were selected for having signs of disturbance in the ionized gas. In our three Seyfert galaxies, the disturbances seem to be related to inflows and outflows in the cold molecular gas, and to the presence of circumnuclear disks.

\subsection{Inflows}\label{sec:disc_inflow}

There is evidence of CO molecular inflows along a bar in NGC\,6860 and along a spiral arm in Mrk\,915. However, for Mrk\,915, the inflow interpretation is only true if we assume a leading spiral arm scenario. Otherwise, it would be consistent with outflows. In these two sources, we might be witnessing the processes responsible for driving material to nuclear regions, which end up feeding the AGN. 
However, the relative importance of different mechanisms in the feeding process is still debatable. Here, we focus on stellar structures associated with inflows in our sources. We refer to \citet{storchi_schnorr19} for a review on the subject. 

Stellar bars -- as in NGC\,6860 -- have long been proposed to play a significant role in feeding AGN activity and also nuclear star formation \citep[e.g.,][]{shlosman+89}. A few studies claim that AGN are more frequently found in barred systems \citep[e.g.,][]{silva-lima+22,alonso+13}, although the majority seem to discard this hypothesis \citep[e.g.,][]{cheung+15,erwin+02,ho+97}. 

Spiral arms -- as in Mrk\,915 --  have been predicted to help drive gas to the central kiloparsec \citep[e.g.,][]{kim_kim14}, with evidence of higher molecular gas concentration in galaxies with more prominent spirals \citep[e.g.,][]{yu+22}. At smaller scales, $\sim$\,10\,--\,100\,pc, there are plenty of examples of inflows along nuclear spirals in AGN \citep[e.g.,][]{bianchin+24,riffelRA+13,riffelRA+08,storchi-bergmann+07}.

Additionally, Mrk\,915 is part of a triple system with signs of a warped outer disk, evidencing tidal interactions with nearby objects (see Figs.\,\ref{fig:companions}). This is one of the several processes that are typically claimed to play a significant role in triggering nuclear activity \citep[e.g.,][]{araujo+23,steffen+23}. It is important to note that not all accreted gas ends up feeding the black hole. For example, the new material might lead to local bursts of star formation \citep{konig+14}.

We note that the three Seyfert galaxies show quite a degree of variability in X-rays and the BLR emission lines. The observed molecular inflows -- in NGC\,6860 and possibly in Mrk\,915 (depending on the disk orientation) -- may be related to this variability. In the sense that episodes of higher accretion rate may both increase the AGN luminosity and partially block the nuclear radiation, leading to a decrease in the luminosity. 
Nonetheless, the molecular inflows observed in our sample are probably only responsible for transporting material to the inner $\sim$\,100\,parsec. And, the short timescales of their X-ray variability are probably associated with episodes of intermittent capture of gas clouds flowing into the central regions \citep{storchi_schnorr19}. 

These CO flows translates to measured cold molecular inflow rate values of $\dot{M}_{\rm{mol,in}}$\,$\sim$\,0.8\,--\,6\,{\msunyr} and 0.09\,--\,0.8\,{\msunyr} for NGC\,6860 and Mrk\,915, respectively. We can compare this with published values in the literature for different objects, to evaluate if these are typical values or outliers.  
The AGN/starburst host Fairall\,49 shows rising $\dot{M}_{\rm{mol,in}}$ values along a bar, reaching $\sim$\,5\,{\msunyr}
\citep{lelli+22}. They note that this value is consistent with the zoom-in cosmological simulations of quasar fueling from \citet{angles-alcazar+21}, although the rate can vary significantly, spanning a range of 0.001\,--\,10\,{\msunyr}. Similarly, \citet{wu+21} found a $\sim$\,12\,{\msunyr} inflow rate in NGC\,3504 along the bar's dust lanes. 
In NGC\,1365, the inflow rates grow from $\sim$\,6\,{\msunyr} in the spirals arms to $\sim$\,40\,{\msunyr} in the bar \citep{elmegreen+09}.
And, in a sample of seven spiral galaxies, most of them barred, \citet{haan+09} observed inflow rates of $\sim$\,0.01\,--\,50\,{\msunyr}. Overall, the inflow rates from our sources do not seem to stand out. 

\subsection{Outflows}\label{sec:disc_outflow}

The three Seyfert galaxies show signs of AGN-induced disturbances in the cold molecular gas, but with their own characteristics. 
In NGC\,6860, the CO outflowing clouds seem to surround an {\oiii} ionized gas outflow. 
In Mrk\,915, the outflow is observed along one of the spirals, although its presence depends on the disk orientation  (scenario of trailing spiral arms). In MCG\,-01-24-012, the disturbance in the CO gas is observed perpendicular to the projected ionization axis, indicating that it propagates equatorially. 

The difference in how the molecular gas is disturbed might be a consequence of different ionization axis inclinations relative to the disk \citep{harrison_ramos-almeida+24}. 
Larger angles between the winds and/or jets axes relative to the galaxy disk result in a weaker coupling between the released energy and the gas in the disk \citep{mukherjee+18}. This is, for example, our proposed scenario for MCG\,-01-24-012 (see Fig.\,\ref{fig:companions}), which might help explain why the spatial anticorrelation between CO and {\oiii} is less pronounced in this source, when compared with our other two Seyfert galaxies \citepalias{dallagnol+25_paperI}. 
We note that the above considerations might not fully apply if the molecular gas originally lies above the disk.

In NGC\,6860 and Mrk\,915, only $f_{\rm{out}}$\,$\sim$\,0.7\,--\,4\,percent of the gas (in mass) is outflowing. In MCG\,-01-24-012, a higher fraction of $\sim$\,30\,percent seems to be disturbed, although the bulk kinematics of the gas in this region follows an ordered motion of rotating in a disk, and the outflow velocity is only $V_{\rm{out}}$\,$\sim$\,40\,\kms.
The corresponding mass outflow rates of our sources are in the range of $\dot{M}_{\rm{mol,out}}$\,$\sim$\,0.09\,--\,3\,{\msunyr}. 
Overall, the $\dot{M}_{\rm{mol,out}}$,  $f_{\rm{out}}$ and outflow velocity ($V_{\rm{out}}$\,$\sim$\,40\,--\,300\,{\kms}) values seems to indicate relatively weak AGN feedback on the cold molecular gas. 
However, as discussed in \citetalias{dallagnol+25_paperI}, part of the molecular gas is likely being depleted by the energy released by the AGN on the ISM. 
This information is not covered by $\dot{M}_{\rm{mol,out}}$ measurements, although it could also be considered a type of negative feedback.

We can compare $\dot{M}_{\rm{mol,out}}$ from our sources with published values from resolved outflow observations.
We can divide the reported $\dot{M}_{\rm{mol,out}}$ measurements in the literature into two classes: weak/moderate cold molecular outflow, in the range  
$\dot{M}_{\rm{mol,out}}$\,$\sim$\,0.1\,--\,10\,{\msunyr}
\citep[e.g.,][]{oosterloo+17,slater+19,alonso-herrero+19,dallagnol+23}; and strong outflows, with values of 10\,--\,10$^2$\,{\msunyr} \citep[e.g.,][]{garcia-burillo+14,alonso-herrero+23,garcia-bernete+21,audibert+19}. Our sample is closer to the first class of weak cold molecular outflows. 
A broader compilation of $\dot{M}_{\rm{out,mol}}$ measurements from the literature, covering 5\,--\,6 orders of magnitude in {\lagn}, have been made by \citet[see their Fig.\,13]{alonso-herrero+23}\footnote{Data from NGC\,7172, measured in that work, and \citet{lutz+20,garcia-burillo+14,ramos-almeida+22,lampert+22}.}, showing a positive trend between both quantities.
In this compilation, there are about eight objects in the same range of luminosities of our three Seyfert galaxies: {\lagn}\,$\sim$\,10$^{43.6}$\,--\,10$^{44.8}$\,{\ergs}. All of the eight objects have $\dot{M}_{\rm{out,mol}}$ values higher than our three sources, by factors of $\sim$\,2\,--\,100 (compared with our maximum value of 3\,{\msunyr}). This indicates that the AGN feedback in NGC\,6860, Mrk\,915, and MCG\,-01-24-012 might be particularly weak, if we assume that our values are not underestimated, or the opposite for the literature values.
However, this is only an exploratory comparison. A more comprehensive comparison should consider uncertainties in the measurements of other authors and try to homogenize the methods used to calculate the mass outflow rates, which are know to vary a lot  \citep[e.g.,][]{harrison+18,dallagnol+21}. 
Such analysis is beyond the scope of this work.
In Section\,\ref{sec:caveats}, we list some sources of uncertainties in our measurements. 

\subsection{Comparing different flow rates}\label{sec:flow}

We can compare the measured $\dot{M}_{\rm{mol,in}}$ with the mass accretion rates required to feed the AGN $\dot{M}_{\rm{acc}}$\,=\,$L_{\rm{AGN}}/(c^2\,\eta)$. 
By assuming an efficiency of conversion of accreted matter into radiation of $\eta$\,=\,0.1 \citep{frank+02}, and using the mean {\lagn} value from Table\,{\ref{tab:sample}}, we obtain $\dot{M}_{\rm{acc}}$ values of $\sim$\,0.009, $\sim$\,0.03, and $\sim$\,0.07\,{\msunyr}, for NGC\,6860, Mrk\,915, and MCG\,-01-24-012, respectively, with a 0.2\,dex uncertainty (from the range in the {\lagn} values). 
Focusing on the sources inflows detections, we obtain ratios of $\dot{M}_{\rm{mol,in}}{/}\dot{M}_{\rm{acc}}$\,$\sim$\,70\,--\,850 (for NGC\,6860) and 3\,--\,40 times (for Mrk\,915, in the leading arms scenario). Hence, the mass inflow rate is larger than the rate required to power the current AGN event in each object. 
It is important to note that comparing $\dot{M}_{\rm{mol,in}}$, at scales of hundreds parsecs, with $\dot{M}_{\rm{acc}}$, which happens at parsec scales, is not straightforward. 
The mechanisms involved in driving gas to nuclear regions, leading to a new AGN event, are probably episodic and occur at different timescales compared to larger-scale inflows \citep{storchi_schnorr19}

We can also compare the mass accretion rate and the molecular outflow rates. 
We obtained ratio values of $\dot{M}_{\rm{mol,out}}/\dot{M}_{\rm{acc}}$\,$\sim$\,9\,--\,140, 3\,--\,30, and 4\,--\,90, for NGC\,6860, Mrk\,915 (in the trailing arms scenario), and MCG\,-01-24-012, respectively.
The accretion rate being lower than the mass outflow rate is in qualitative agreement with what has been observed for the ionized and hot molecular gas \citep[e.g.,][]{muller-sanchez+11, bianchin+22, barbosa+09, riffelRA_storchi-bergman11,schnorr-muller+14}. If we assume that the AGN feedback is only responsible for the observed disturbances in the CO, this suggests that only a fraction of the outflowing clouds are launched directly by the energy released by the AGN. The remaining outflowing gas might originate from the surrounding ISM, which could have been disturbed and pushed away in the past by the AGN events. Moreover, part of the ejected molecular gas could have been driven by local starbursts. 
 We note, however, that there are no clear signs of a young stellar population close to the disturbed regions in NGC\,6860 \citep{lipari+93,bennert+06} and in Mrk\,915 \citep{trippe+10,bennert+06}, as judged from emission line ratios and optical continuum.
Mrk\,915 seems to have an excess of UV continuum emission, which could be a hint of a nuclear starburst \citep{munoz-marin+09}, but a careful modeling of the UV continuum is needed to discard other contributions, like the AGN continuum, scattered light, and nebular emission.
For MCG\,-01-24-012, we did not find external evidence in favor (or not) of ongoing star formation in the double-peak CO region.

Finally, we can compare the mass inflow and outflow rates. 
For Mrk\,915, the inflow and outflow scenarios are mutually exclusive, making a comparison between them not possible.   
We, therefore, will focus only in NGC\,6860, where the ratio between these values is $\dot{M}_{\rm{mol,out}}/\dot{M}_{\rm{mol,in}}$\,$\sim$\,0.1. And comparing all flow rates, we have that $\dot{M}_{\rm{acc}}$\,<\,$\dot{M}_{\rm{mol,out}}$\,$\lesssim$\,$\dot{M}_{\rm{mol,in}}$ in this source. 
Analogously to the comparison between $\dot{M}_{\rm{mol,in}}$ and $\dot{M}_{\rm{acc}}$, comparing the $\dot{M}_{\rm{mol,out}}$ and $\dot{M}_{\rm{mol,in}}$ is not straightforward due to different spatial and temporal scales involved. If we ignore that, the above ratio suggests that part of the infalling CO clouds might be ejected and/or disturbed by the AGN feedback before reaching the nuclear region.  
We note that this considers only the cold molecular phase traced by CO(2-1) and the impact on it associated with the AGN feedback.

The above discussions raise the question of whether the inflowing cold molecular gas leads to an accumulation of a nuclear reservoir of molecular gas. 
Besides the fraction that is observed in outflow, part of it might be depleted by the ionized outflow, jets, and/or radiation, as suggested by spatial anticorrelation between CO and the {\oiii} ionized gas emission within the inner kiloparsec region of these two sources \citep{dallagnol+25_paperI}. 
We believe that we are observing a scenario analogous to the results obtained by \citet{garcia-burillo+24}. They found evidence that the molecular reservoir builds up over time in the central regions, with higher molecular gas concentrations being observed in more luminous AGN, probably a consequence of higher AGN mass accretion rates. For objects with X-ray luminosities above $\sim$\,$10^{41.5}$\,{\ergs} -- as our sources -- the negative AGN feedback starts to affect the molecular gas nuclear reservoir, decreasing the nuclear concentration. 
In addition, part of the extra fuel can also be used to form new stars. However, as discussed above, we did not find strong evidence for ongoing nuclear star formation.

\subsection{Caveats}\label{sec:caveats}

Here we list some caveats of our results.
Probably the highest source of uncertainty for mass-related measurements is the $r_{21}$ and {\alphaCO} factors, used to convert the CO(2-1) flux to H$_2$ masses (see Appendix\,\ref{sec:inflow_outflow}, for details), and for which we do not have direct measurements for our sources. 
Together, the wide range of $r_{21}$\,=\,0.8\,--\,1.2 and {\alphaCO}\,=\,0.8\,--\,4.3\,{\msunKkmspc} values used in this work introduce an order-of-magnitude uncertainty in molecular mass values. 
We choose to use such ranges because they cover different physical scenarios. 
In particular for outflowing CO clouds, {\alphaCO} values closer to our minimum range have been used in the literature \citep[e.g.,][]{morganti+15,dasyra+16,ramos-almeida+22}. 
In addition, direct measurements, using CO rovibrational infrared lines, indicate that the {\alphaCO} might be even smaller by a factor of up to $\sim$2 \citep{pereira-santaella+24}. In such a case, when dealing with the values presented in this work, one should consider the lower values of the $M_{\rm{mol,out}}$ and $\dot{M}_{\rm{mol,out}}$ ranges as more reliable measurements.

For simplicity, we assumed that both the inflows and outflows occur at the same plane, with an inclination obtained for the gas rotating in a disk. 
However, when inflowing and outflowing CO clouds are observed along the same line of sight, it is reasonable to suspect that the outflow occurs outside the disk plane. 
For example, assuming an outflow above the disk in NGC\,6860 (inclination > 57.7{\degree}), would increase the resulting $R_{\rm{out}}$ values while decreasing $V_{\rm{out}}$ and $\dot{M}_{\rm{mol,out}}$. 
By varying the inclinations by $\Delta i$\,$\sim$\,10\,--\,15{\degree}, we showed that values of $\dot{M}_{\rm{mol,out}}$ (and $\dot{M}_{\rm{mol,in}}$) can vary by a factor of up to $\sim$\,3 (see Sects\,\ref{sec:ngc6860_res_bar_inflow}, \ref{sec:ngc6860_res_outflow}, \ref{sec:inflow_mrk915_leading} and \ref{sec:mcg0124012_res_doublepeak}).

Another source of uncertainty is the definition of outflow velocity, here we calculated from the LoS velocity residuals of the CO component tracing the outflow. Other authors have used different definitions that return higher values of $V_{\rm{out}}$ \citep[e.g.,][]{sun+17,karouzos+16,bischetti+19}. 
For example, using $V_{\rm{out}}\,{=}\,\vel\,{+}\,2\,\sigma$ increases $M_{\rm{mol,out}}$ by a factor of up to $\sim$\,3 in our sources, which is still within the uncertainty associated with the mass. 
We refer to \citet{harrison+18,harrison_ramos-almeida+24} for a broader discussion on the different methods and assumptions, like the outflow geometry, that one can use to calculate mass outflow rates.

Inflows and outflows were identified based on the residuals of the 2D rotating disk models. We have tested different scenarios in NGC\,6860 and MCG\,-01-24-012, and considered the two possible disk orientations for Mrk\,915. However, from our analysis, we cannot discard that more complex models might lead to different interpretations of the results.

\section{Conclusions}\label{sec:conclusions}

We analyzed the cold molecular gas distribution and kinematics -- as traced by CO(2-1) -- in three local Seyfert galaxies observed with ALMA. The high angular resolution of the ALMA observations allowed us to resolve the gas on scales of $\sim$\,0.5\,--\,0.8{\arcsec} ($\sim$\,150\,--\,400\,pc). 
Below, we summarize the picture drawn for each object.

    \paragraph{NGC\,6860.} The CO emission is observed along a stellar ring and its internal stellar bar. Along the ring, the molecular gas mostly follows the disk rotation. 
    We discussed two scenarios for CO in the bar: a molecular gas inflow along the bar with an inflow rate of $\dot{M}_{\rm{mol,in}}$\,$\sim$\,0.8\,--\,6\,\msunyr, or a decoupled circumnuclear disk (CND), rotating in a different plane from that of the gas in the ring. 
    There is also a molecular gas outflow within $R_{\rm{out}}$\,$\sim$\,560\,pc from the nucleus, surrounding part of the bipolar {\oiii} ionization cone and outflow, and with a mass outflow rate of $\dot{M}_{\rm{mol,out}}$\,$\sim$\,0.1\,--\,1\,\msunyr. 
    
    \paragraph{Mrk\,915.} The cold molecular gas mostly lies along spiral arms, with rotation dominating the kinematics. 
    There is a region with CO double-peak profiles along one of the spiral arms, indicating the presence of two kinematic components. 
    Due to the uncertain orientation of the galaxy's disk, one of the components can be either tracing an inflowing stream of molecular gas or an outflow, with the galaxy having leading or trailing spiral arms, respectively. In the hypothesis of inflows, the mass inflow rate would be with $\dot{M}_{\rm{mol,in}}$\,$\sim$\,0.09\,--\,0.8\,\msunyr,
    while in the case of outflow, the mass outflow rate would be $\dot{M}_{\rm{mol,out}}$\,$\sim$\,0.09\,--\,0.7\,\msunyr with $R_{\rm{out}}$\,$\sim$\,2.8\,kpc.
    The influence of nearby galaxies might play a role in both scenarios.
    We note that the lower S/N of Mrk\,915's spectra might be affecting our conclusions. Deeper observations are needed to better model its kinematics.

    \paragraph{MCG\,-01-24-012.} The cold molecular gas is partly located along a $\sim$\,3\,kpc (radius) stellar ring, with kinematics dominated by rotation. CO emission is also observed inside the ring, including along the stellar bar. 
    In the inner kiloparsecs, the CO emission extends almost perpendicularly to the {\oiii} ionization cone, with an orientation distinct from the outer galaxy disk. 
    We identify this region as a possible circumnuclear disk, with molecular gas outflowing along it, but with the kinematics still being dominated by rotation. 
    This effect is probably caused by an ionized gas outflow associated with the bipolar {\oiii} emission that disturbs the molecular gas and produces an equatorial outflow with $\dot{M}_{\rm{mol,in}}$\,$\sim$\,0.4\,--\,3\,\msunyr and $R_{\rm{out}}$\,$\sim$\,3\,kpc.


Analyzing the results of the three Seyfert galaxies together, we find the following common features:  
\begin{itemize}

    \item The bulk of molecular gas is rotating in the galaxy disk, near stellar structures identified as bars, rings, and spirals. 
    Within $\sim$\,0.56\,--\,3\,kpc, we found disturbances in the molecular gas, as traced by CO double-peaks spectral profiles. We showed that these disturbances are associated with CO inflows and outflows, and that the gas might be rotating in a different plane as part of a circumnuclear disk, as in the case of MCG\,-01-24-012, and possibly in NGC\,6860.

    \item 
    The mass outflow rates are in the range $\dot{M}_{\rm{mol,out}}$\,$\sim$\,$0.09$\,--\,$3$\,\msunyr. This suggests a particular weak negative AGN feedback in the cold molecular gas, when compared with other sources with similar AGN luminosities, {\lagn}\,$\sim$\,$10^{44}$\,{\ergs}( $\dot{M}_{\rm{mol,out}}$ values larger by factors of $\sim$\,2\,--\,100).
    
    \item Only a fraction of the inflowing molecular clouds end up feeding the AGN since the inflow rates of NGC\,6860 and Mrk\,915 (in the trailing arms scenario) are $\sim$\,3\,--\,850 times the AGN mass accretion rate. 
    
    \item 
    The AGN accretion rate is also not enough to feed the observed CO outflow mass rates, being lower by factors of $\sim$\,3\,--\,140. 
    This indicates that the mass outflow rate is due to mass-loading of nuclear outflows from molecular gas that has accumulated over time in the circumnuclear region. There is also a possible contribution of nuclear starbursts, although we did not find clear evidence of it in our sources.
    
\end{itemize}

Although the measured $\dot{M}_{\rm{mol,out}}$ indicates a weak impact in this gas phase, the presence of regions with a CO deficit in our three sources suggests that the AGN feedback might be depleting part of the molecular gas content \citepalias{dallagnol+25_paperI}. This is an impact that is not taken into account by $\dot{M}_{\rm{mol,out}}$, and should receive more attention in future works.
In agreement with results from \citet{garcia-burillo+24}, our sources might be increasing the CO reservoir close to the nucleus, given the measured inflow rates. This build-up then leads to an increase in the AGN mass accretion rate, which increases the AGN total energy output. Given the current AGN luminosity, we might be observing the beginning of the depletion of the nuclear molecular reservoir in our sources.

\begin{acknowledgements}

This study was financed in part by the Coordena{\c c}{\~a}o de Aperfei{\c c}oamento de Pessoal de N\'ivel Superior (CAPES-Brasil, 88887.478902/2020-00, 88887.985730/2024-00). 
RAR acknowledges the support from Conselho Nacional de Desenvolvimento Cient\'ifico e Tecnol\'ogico (CNPq; Proj. 303450/2022-3, 403398/2023-1, \& 441722/2023-7), Funda\c c{\~a}o de Amparo \`a pesquisa do Estado do Rio Grande do Sul (FAPERGS; Proj. 21/2551-0002018-0), and CAPES (Proj. 88887.894973/2023-00).\\

This paper makes use of the following ALMA data: ADS/JAO.ALMA\#2018.1.00211.S. ALMA is a partnership of ESO (representing its member states), NSF (USA) and NINS (Japan), together with NRC (Canada), MOST and ASIAA (Taiwan), and KASI (Republic of Korea), in cooperation with the Republic of Chile. The Joint ALMA Observatory is operated by ESO, AUI/NRAO and NAOJ.\\

The Legacy Surveys consist of three individual and complementary projects: the Dark Energy Camera Legacy Survey (DECaLS; Proposal ID \#2014B-0404; PIs: David Schlegel and Arjun Dey), the Beijing-Arizona Sky Survey (BASS; NOAO Prop. ID \#2015A-0801; PIs: Zhou Xu and Xiaohui Fan), and the Mayall z-band Legacy Survey (MzLS; Prop. ID \#2016A-0453; PI: Arjun Dey). DECaLS, BASS and MzLS together include data obtained, respectively, at the Blanco telescope, Cerro Tololo Inter-American Observatory, NSF’s NOIRLab; the Bok telescope, Steward Observatory, University of Arizona; and the Mayall telescope, Kitt Peak National Observatory, NOIRLab. Pipeline processing and analyzes of the data were supported by NOIRLab and the Lawrence Berkeley National Laboratory (LBNL). The Legacy Surveys project is honored to be permitted to conduct astronomical research on Iolkam Du’ag (Kitt Peak), a mountain with particular significance to the Tohono O’odham Nation.\\

This research has made use of the NASA/IPAC Extragalactic Database, which is funded by the National Aeronautics and Space Administration and operated by the California Institute of Technology.\\

This work used the color scheme from \citet{wong+11} in some of the plots to minimize confusion for colorblind viewers.
Figure\,\ref{fig:models} was generated by combining the following software: {\sc FreeCad}, {\sc LibreOffice Draw} and {\sc Matplotlib}.

\end{acknowledgements}


\bibpunct{(}{)}{;}{a}{}{,}
\bibliographystyle{aa_url}
\bibliography{bib}


\begin{appendix}

\section{Grid of spectra}\label{sec:grid}

\begin{figure*}[!ht]
     \centering
    \begin{subfigure}[b]{0.346\textwidth}
        \centering
    	\includegraphics[width=1\linewidth]{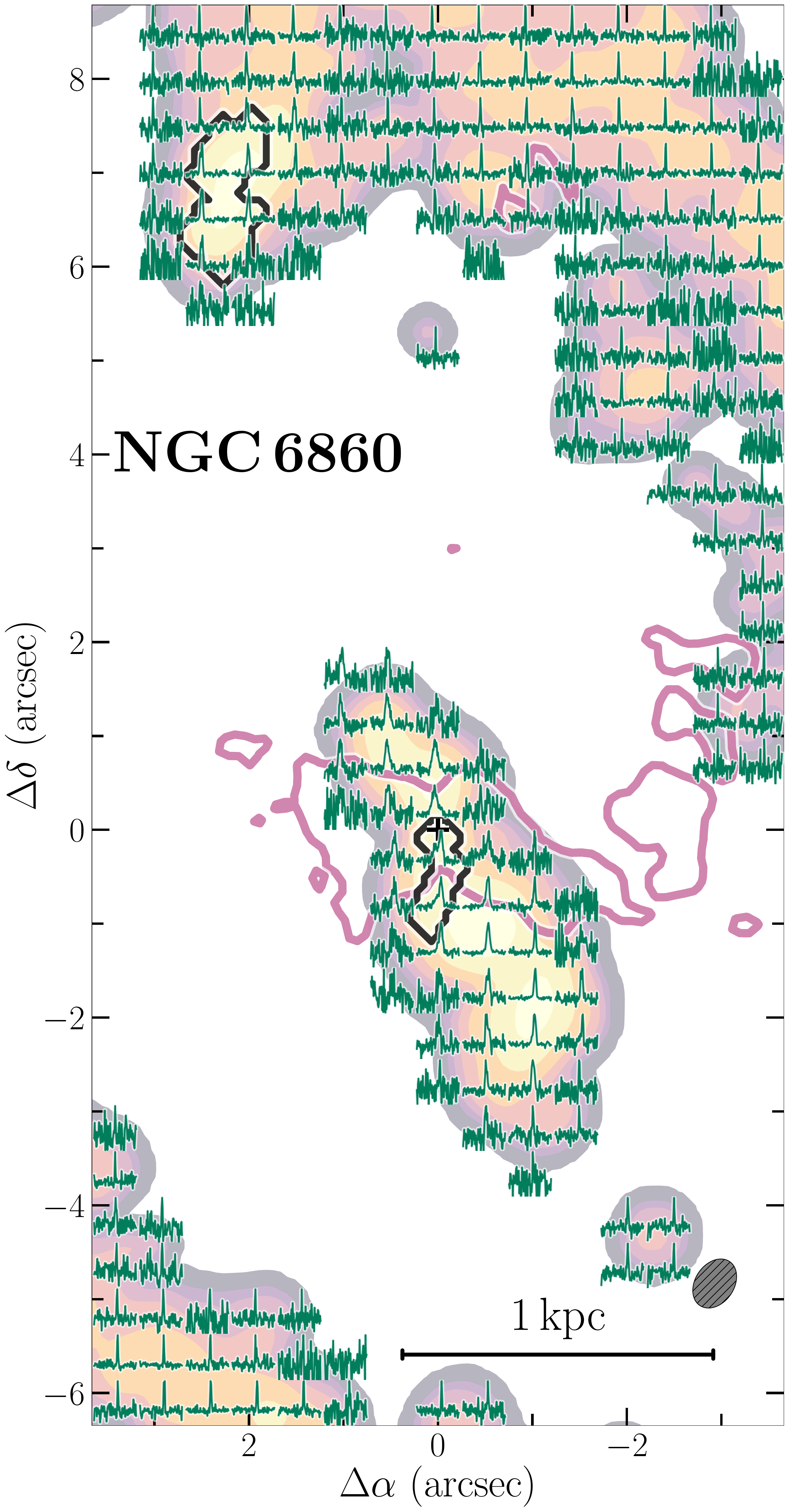}
    \end{subfigure}
    \begin{subfigure}[b]{0.3075\textwidth}
        \centering
    	\includegraphics[width=1\linewidth]{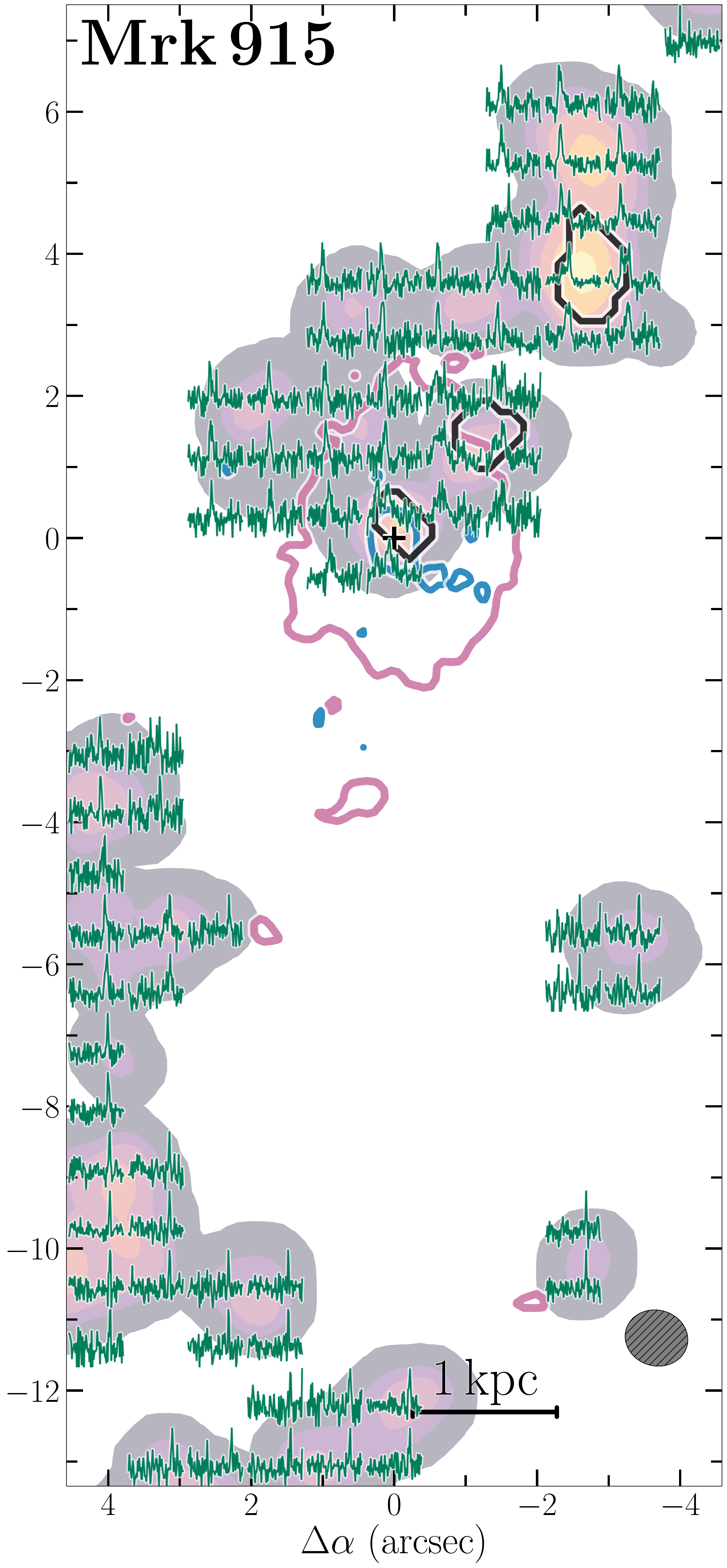}
    \end{subfigure}
    \begin{subfigure}[b]{0.3307\textwidth}
        \centering
    \includegraphics[width=1\linewidth]{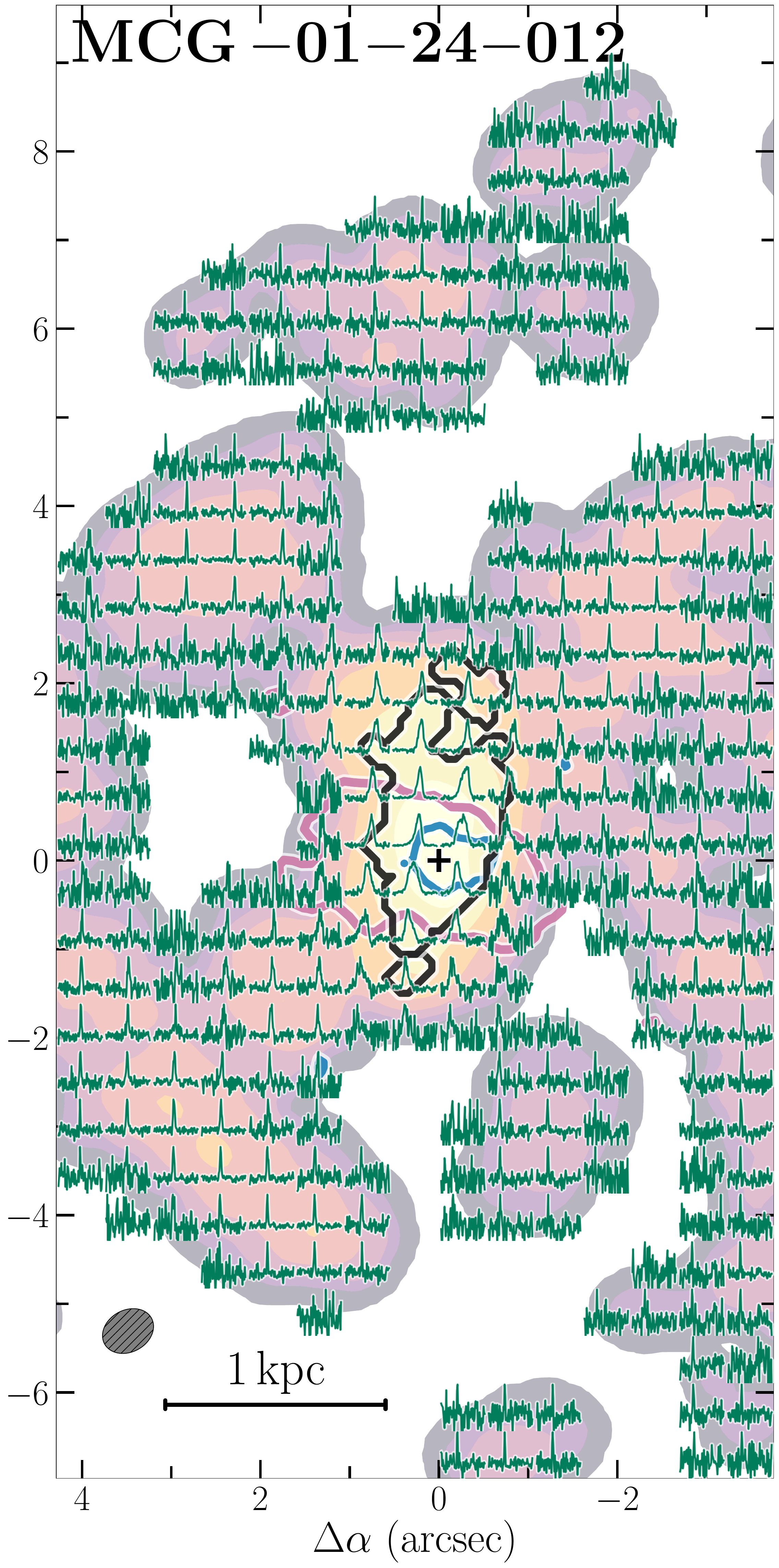}
    \end{subfigure}   
\caption{Grid of the spectra showing the spatial variation of CO(2-1) emission-line profiles of NGC\,6860, Mrk\,915, and MCG-01-24-012. Each grid spectrum is the sum of all individual spectra inside a region equivalent to the observation beam (gray ellipses). All spectra are normalized for better visualization. 
The velocity range of the profiles is (-400,\,400)\,{\kms}.
For reference, the CO flux distribution is shown in the translucent colormap. Overlaid are outer contours from the VLA 3.6\,cm radio (in blue) and the smoothed {\hst} {\oiii} emission (in reddish-purple). 
We note that the measurements presented on Tables\,\ref{tab:inflow} and \ref{tab:outflow} were calculated from individual spaxels, and not from the integrated spectra presented in these grids.}
\label{fig:grids}
\end{figure*}

To help visualize the variation of spectral CO(2-1) profiles over the FoV, we generated a grid of spectra, displayed in Fig.\,\ref{fig:grids}. 
Each grid spectrum corresponds to the sum of all individual spectra inside square regions, with sides equal to the average between the minor and major axes of FWHM$_{\rm{beam}}$: $\sim$\,0.5\,--\,0.8{\arcsec}, corresponding to $\sim$\,150\,--\,400\,pc.

\section{Moments and mask}\label{ap:mom-mask}

To calculate the moments, we used only pixels that were not masked. These 3D masks were generated reproducing a procedure described in \citet{diteodoro_fraternali15}: first, the original data cubes were smoothed, channel by channel, by convolution with their beam multiplied by a factor of $k_{\rm{mult}}$ = 2; then, any resulting values below a $k_{\rm{cut}}$ = 2\,{\sigmarms} cut were masked. For Mrk\,915, we used values of $k_{\rm{mult}}$ = 3 and $k_{\rm{mult}}$ = 1.
Figures\,\ref{fig:moments_6860}, \ref{fig:moments_mrk915} and \ref{fig:moments_mcg0124102} show that $M_1$ and $M_2$ maps are less extended relative to $M_0$. This happens because we added another constraint: we masked pixels with flux density below 3\,$\sigma$ in the non-convolved data. This was necessary since these $M_1$ and $M_2$ are more affected by pixels with low S/N.
The 3D mask was only used to obtain the moments, but a spectrally collapsed 2D version was used in the other analysis as the limit of the CO emission. 

\section{Leading spiral arms in Mrk\,915?}\label{ap:mrk915-orientation}

\begin{figure*}
    \includegraphics[width=1\linewidth]{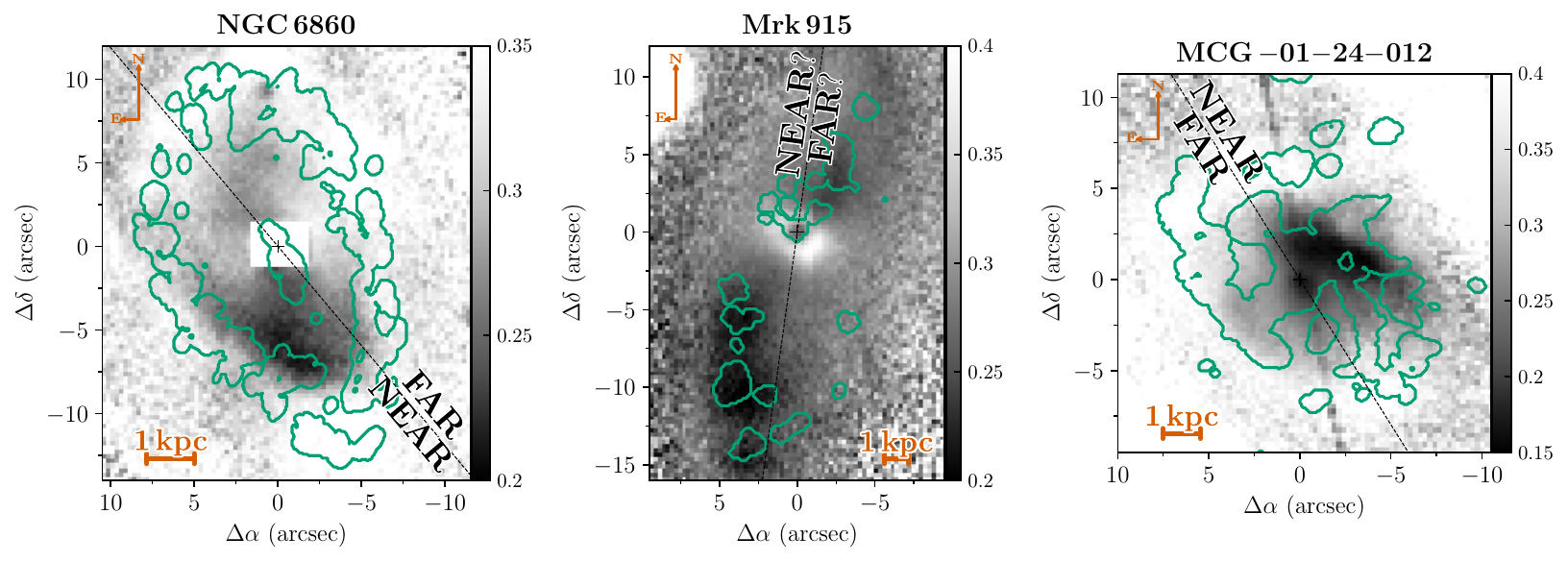}
    \caption{Maps of the ratio between images obtained at the g and z-bands (from Dark Energy Camera), used as a proxy for dust attenuation. The CO(2-1) $M_0$ outer contours are overlaid in blueish-green color. The near side of the disk is the southeast and the northwest in NGC\,6860, and MCG\,-01-012-24, respectively.
    For  Mrk\,915, the g / z map (middle panel) suggests that the near side of the disk could be in the east, but this would result in Mrk\,915 having a rare case of leading arms (see discussion in Sect.\,\ref{ap:mrk915-orientation}).}
    \label{fig:gz}
\end{figure*}

Figure\,\ref{fig:gz} displays the maps of the ratio between images in g and z-bands taken with filters of the Dark Energy Camera. 
As described in Sect.\,\ref{sec:dust}, the g / z ratio map was used as a proxy for dust attenuation and to identify the near side of the galaxy disk. More attenuated regions (lower g / z ratio values) are expected to be observed at the near side of the disk. 
In particular, for Mrk\,915, the map presents lower values at the southeast of the nucleus, indicating that the near side of the disk is to the southeast. 
However, if the stars and CO co-rotate in putative trailing arms, from the CO(2-1) $M_1$ velocity field (Fig.\,\ref{fig:moments_mrk915}), the near side should be on the west: a contradiction.
Is Mrk\,915 a rare case of a galaxy with leading arms?

Part of the confusion might originate from the spiral arms and dust lanes being mostly close and along the galaxy’s photometric major axis (instead of the minor axis), making the contrast with the bulge's light more subtle. Consequently, using dust attenuation to discover the orientation of the disk in Mrk\,915 might lead to erroneous conclusions. 
This strategy also depends on the Mrk\,915's optical continuum having a significant contribution from the bulge's stellar light, which we could not confirm due to the lack of a bulge-to-disk decomposition in the literature for this object.

If we reverse the orientation -- assuming that the near side is to the west -- then the arms would be trailing. In this case, the dynamics of the  Mrk\,915's spiral arms would follow the expected from most theoretical models \citep[e.g.,][]{evans_read98} and simulations \citep[e.g.,][]{sellwood_athnassoula86}, where leading arms are expected to be more unstable, with a posterior decay to a trailing arms motion \citep{toomre81}. But we note that long-living leading arms have also been proposed \citep[e.g.,][]{lieb+22}. On the observational side, \cite{iye+19} found that all objects in a sample of 146 galaxies possess trailing arms, with the authors using the prevalence of dust lanes along the minor axis to identify the near side of the galaxy disk, reproducing the results found by \citet{deVaucouleurs58}. A rare case of a galaxy with evidence of leading arms is NGC\,4622 \citep{byrd+08}. In that work, the authors suggest that the phenomenon could be a consequence of an encounter with a companion, as reproduced in simulations \citep[e.g.,][]{byrd+93}, which is a scenario that cannot be discarded in Mrk\,915 due to the presence of nearby objects (see Fig.\,\ref{fig:companions}). 
Other proposed examples are ESO\,297-27 \citep{grouchy+08}, and IRAS\,18293-3134 \citep{vaisanen+08}, both with nearby companions, with the latter one being more similar to Mrk\,915 given the absence of outer spirals wound up in the opposite direction from the inner ones.

Since we cannot be sure of the true orientation, we will use the observation of stronger dust attenuation to identify the near side of the disk as the default in the figures (near side in the east). Nevertheless, we will also consider the opposite orientation in the analysis/discussion of this galaxy.


\section{Disk model fit}\label{ap:bertola}

We fitted the LoS velocity distributions ($\vel_{\tt{c1}}$) 
using a disk model with circular orbits with velocity (V$_{\rm{c}}$) of the form  \citep{bertola+91}:
\begin{equation}
    V_{\rm{c}} (r) = \frac{r\, V_{\rm{max}} }{(r^2 + c_0^2)^{p/2}},
\end{equation}
with V$_{\rm{max}}$ being the maximum circular velocity, and $p$ a parameter that governs the shape of the radial velocity profile. 
For the fit, we also consider that the observed LoS velocity depends on the inclination ($i$) of the disk, the position angle (PA) of the kinematic major axis, and the systemic velocity of the host galaxy $\vel_{\rm{sys}}$. 
Following  \citet{bertola+91}, we let $p$ range between 1 (asymptotically flat) and 3/2 (finite total mass). The spatial scale parameter $c_0$ is such that, in $r = c_0$  for $p=1$, the circular velocity is  V$_{\rm{c}} (r=c_0) \sim 0.7 \vel_{\rm{max}}$.
Note that the units of the denominator and the ``$r$'' in the numerator only cancel out when p\,=\,1. Otherwise, there is an excess of $[r]^{1-p}$, where $[r]$ is the unit of distance used during the fit ($[r]$\,=\,{\arcsec}, in our case). We emphasize that the above velocity function was used to model the velocity field, in order to analyze the velocity residuals, and, therefore, it should ultimately be considered only an analytical model.

The kinematic model center was fixed at the ALMA millimeter continuum peak (see Table\,\ref{tab:sample}). 

Fig.\,\ref{fig:bertola} also shows the best model and the residuals (for both components), while Table\,\ref{tab:bertola2} lists the best-fit parameter values. 
The values and uncertainty are the mean and standard deviation, resulting from 100 fits done on the original velocity field -- the masked $\vel_{{c1}}$ map (in {\kms}) -- added by a Gaussian random noise. This noise map was scaled by the cube channel width (in {\kms}), which is of the order of the spectral resolution, although larger since we re-binned the data cubes (see Sect.\,\ref{sec:sample_observations}). Each fit is done by minimizing the residuals.

\section{Mass outflow and inflow rate}
\label{sec:inflow_outflow}

Here, we describe how the outflow and inflow-related quantities were calculated and later displayed in Tables\,\,\ref{tab:inflow} and \ref{tab:outflow}. 

The cold molecular gas masses ($M_{\rm{mol}}$) were calculated from the velocity-integrated fluxes ($S_{\nu} \Delta \vel_{\rm{CO(2-1)}}$) using the same method detailed in \citetalias{dallagnol+25_paperI}. In brief, in each spaxel, $S_{\nu} \Delta \vel_{\rm{CO(2-1)}}$ was derived from the Gaussian profiles fitted to the emission lines (Sect.\,\ref{sec:line-fitting}), and then converted to CO(2-1) luminosities \citep[$L'_{\rm{CO(2-1)}}$,][]{solomon+97}. The CO(1-0) luminosity was obtained after assuming a ratio of $r_{21}$\,=\,$L'_{\rm{CO(2-1)}}/L'_{\rm{CO(1-0)}}$\,=\,0.8\,--\,1.2, typical of nearby galaxies \citep{braine_combes92}, Finally, $M_{\rm{mol}}$ are estimated using a conversion factor of 
{\alphaCO}\,=\,$M_{\rm{H_2}}/L'_{\rm{CO(1-0)}}$\,=\,0.8\,--\,4.3\,{\msunKkmspc}, covering a wide range of physical conditions for the clouds, with the {\alphaCO} limits corresponding to average values obtained for in ultra luminous infrared galaxies and in the Milk Way \citep{bolatto+13}.

\subsection{Mass inflow rate}\label{ap:inflow}

Based on \citet{storchi-bergmann+07,riffelRA+08}, we calculated the mass inflow rate by assuming a constant flow over a cylindrical geometry $\dot{M}_{\rm{mol}}=\rho_{\rm{mol}}V_{\rm{in}}\it{A}$,    
where $V_{\rm{in}}$ is the deprojected inflow velocity, $A$ is the cross-section area of an elliptic cylinder, and $\rho_{\rm{mol}}$ is the volumetric mass density of the cold molecular gas inside the cylinder. 
If we calculate the density inside the cylinder from $\rho_{\rm{mol}}$\,=\,$M_{\rm{mol,in}} / (h \,A)$, the mass inflow rate became simply: 
\begin{equation}
\dot{M}_{\rm{mol,in}} = \frac{M_{\rm{mol,in}}V_{\rm{in}}}{h},\label{eq:inflow}
\end{equation}
where is $h$ is the projected length of the cylinder, $M_{\rm{mol,in}}$ is the molecular mass inside it and  $V_{\rm{in}}$ is the de-projected inflow absolute velocity.

We can also estimate the average volumetric mass density in the inflows using $\rho_{\rm{mol}}$\,=\,$M_{\rm{mol,in}}/(h \pi r_a\,r_d)$, where $r_{a}$ is the deprojected major radii of cylinders faces and $2\,r_d$ is the disk height, assumed to be $2\,r_d$\,$\sim$\,30\,pc \citep{hicks+09}.
In addition, the volumetric number density of molecular clouds can be calculated from  $n_{\rm{mol}}$\,=\,$\rho_{\rm{mol}}/(1.36\cdot2\,m_p)$\,$\sim$\,4\,--\,30\,cm$^{-3}$, where $m_p$ is the proton mass, and 1.36 factor accounts for 36\,\% contribution from Helium \citep{saintonge+22}.

\begin{figure}
	\includegraphics[width=1\linewidth]{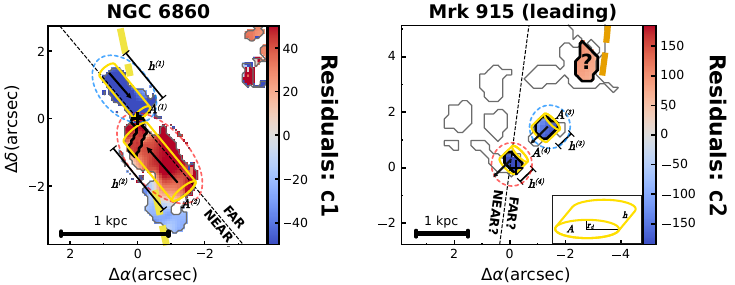}
    \caption{
    Geometry of the molecular inflow along the bar (dashed yellow line) in NGC\,6860, and along one of the spiral arms in Mrk\,915 (assuming that the arms are leading). The yellow elliptic cylinders have heights $h$ and areas $A=\pi r_a r_d$, where $r_d$ is half the height of the disk and $r_a$ is the major radius of the face (see inset in the right panel). The superscript numbers identify the regions used in calculations in Sect.\,\ref{ap:inflow}. 
    The cylinders are drawn over the LoS residuals of the disk model (as Fig.\,\ref{fig:bertola}). These residuals associated with components used in the $\dot{M}_{\rm{mol,in}}$ calculations: \texttt{c1} for NGC\,6860 and \texttt{c2} for Mrk\,915. The reddish-purple and blue dashed ellipses delimit the regions used to obtain the velocities and masses of the inflows coming from the north and the south in NGC\,6860. For Mrk\,915, they mark the region of the farthest and the nearest inflowing clumps. 
    }
    \label{fig:inflow}
\end{figure}

\paragraph{\textbf{NGC\,6860.}} 
Using Eq.\,\ref{eq:inflow} and assuming a constant inflow rate inside the cylindrical geometry (sketch in left panel of Fig.\,\ref{fig:inflow}) we found cold molecular mass inflow rates of:  $\dot{M}_{\rm{mol,in}}^{(1)}$\,$\sim$\,0.34\,--\,2.7\,{\msunyr}, from the north of the nucleus, at the far side of the disk; 
and $\dot{M}_{\rm{mol,in}}^{(2)}$\,$\sim$\,0.46\,--\,3.7\,{\msunyr}, from the south. And integrating from both directions: $\dot{M}_{\rm{mol,in}}$\,$\sim$\,0.8\,--\,6\,{\msunyr} (see Table\,\ref{tab:inflow}), with the large uncertainty being dominated by the uncertainty in the {\alphaCO} value. 
These values are obtained assuming an inclination of $i$\,=\,57.7{\degree}, for an inflow co-planar with the disk. 
In the above calculations, we used as the de-projected inflow velocities values: $V_{\rm{in}}^{(1)}$\,$\sim$\,$-90$\,{\kms} and $V_{\rm{in}}^{(2)}$\,$\sim$\,90\,{\kms} (the superscript number matches the numeration in Fig.\,\ref{fig:inflow}). These values correspond to the average of the LoS residuals -- relative to the disk model --  of the $c_1$-component inside the dashed ellipses of Fig.\,\ref{fig:inflow}. After integrating the flux inside these ellipses, we obtained molecular masses of $M_{\rm{mol,in}}^{(1)}$\,$\sim$\,(0.34\,--\,2.6)\,${\times}$\,10$^{7}${\msun} and $M_{\rm{mol,in}}^{(2)}$\,$\sim$\,(0.8\,--\,6.5)\,${\times}$\,10$^{7}${\msun} in each region,  resulting in a total of $M_{\rm{mol,in}}$\,$\sim$\,(1\,--\,9)\,${\times}$\,10$^{7}${\msun}.
The deprojected cylinder lengths used in the calculation are $h^{(1)}$\,$\sim$\,1.6{\arcsec}$\sim$\,900\,pc and $h^{(2)}$\,$\sim$\,2.8{\arcsec}$\sim$\,1.6\,kpc.

Using $r_{a}^{(1)}$\,$\sim$\,0.29\arcsec\,$\sim$\,160\,pc and $r_{a}^{(2)}$\,$\sim$\,0.42\arcsec\,$\sim$\,240\,pc as values for deprojected major radii of cylinders' faces (see Fig.\ref{fig:inflow}), we obtained a volumetric mass density in the inflows of $\rho_{\rm{mol}}$\,=\,$\sim$\,0.2\,--\,2\,{\msun}\,pc$^{-3}$. Here, $\rho_{\rm{mol}}$ is the average of the values obtained for the north and the south inflows, which are approximately equal. 
This corresponds to a volumetric number density of molecular clouds of $n_{\rm{mol}}$\,$\sim$\,3\,--\,30\,cm$^{-3}$.

\paragraph{\textbf{Mrk\,915 (leading arms scenario).}}
Applying the Eq.\,\ref{eq:inflow} individually for each inflowing CO clump of Mrk\,915 (see sketch in Fig.\,\ref{fig:inflow}), we obtained mass inflow rates of: 
$\dot{M}_{\rm{mol,in}}$\,$\sim$\,$\dot{M}_{\rm{mol,in}}^{(3)}$\,$\sim$\,$\dot{M}_{\rm{mol,in}}^{(4)}$\,$\sim$\,0.09\,--\,0.8\,{\msunyr}. Since the inflows come from the same direction, we average the mass inflow rates from both clumps, which have essentially the same value.
The inflow de-projected velocities of each clump are $V_{\rm{in}}^{(3)}$\,$\sim$\,$-100$\,{\kms} and $V_{\rm{in}}^{(4)}$\,$\sim$\,$-160$\,{\kms}, while the molecular masses are 
$M_{\rm{mol,in}}^{(3)}$\,$\sim$\,(0.35\,--\,3.0)\,${\times}$\,10$^{6}${\msun} and $M_{\rm{mol,in}}^{(4)}$\,$\sim$\,(0.56\,--\,4.8)\,${\times}$\,10$^{6}${\msun}. 
Here, we assumed an inflow inclination of $i$\,=\,66.5{\degree} (co-planar with the disk). 

For each clump, we considered the values of \texttt{c2}-component inside the blue (farthest clump) and reddish-purple (nearest) dashed ellipses shown in Fig.\,\ref{fig:inflow}.
We assumed the same geometry for both cylinders: $h^{(3)}$\,=\,$h^{(4)}$\,=\,$2r_{a}^{(3)}$\,=\,$2r_{b}^{(4)}$\,$\sim$\,0.5{\arcsec}$\sim$\,600\,pc, using an inclination of $i$\,=\,66.5{\degree}. The average volumetric mass and number densities are $\rho_{\rm{mol}}$\,$\sim$\,0.03\,--\,0.2\,{\msun}\,pc$^{-3}$ and 
$n_{\rm{mol}}$\,$\sim$\,0.4\,--\,3\,cm$^{-3}$, respectively, with the densities being $\sim$\,2 times higher in the farthest clump. 

\subsection{Mass outflow rate}\label{ap:outflow}

By assuming that the mass outflow rate is constant over time, we can calculate it using \citep{lutz+20}:
\begin{equation}
\dot{M}_{\rm{mol,out}}=\frac{M_{\rm{mol,out}}\,V_{\rm{out}}}{R_{\rm{out}}},\label{eq:outflow}
\end{equation}
where $M_{\rm{mol,out}}$ is the total outflowing gas mass, and $V_{\rm{out}}$ and $R_{\rm{out}}$ are the de-projected maximum outflow velocity and extent, respectively.

\paragraph{\textbf{NGC\,6860.}} 
The total molecular mass of the outflowing clouds in NGC\,6860 is $M_{\rm{mol,out}}$\,$\sim$\,$0.6$\,--\,$5$\,$\times$\,$10^6$\,{\msun}, as obtained from the flux of the component \texttt{c2} of  $S_{\nu} \Delta \vel_{\rm{CO(2-1),out}}$\,$=$\,340$\pm$6\,{\mjykms}, integrated inside a 0.6{\arcsec} radius (see Table\,\ref{tab:outflow}). This corresponds to $\sim$\,0.7 percent of the total molecular mass in the galaxy observed within a radius of $\sim$\,13{\arcsec}.
Assuming that the outflow is almost co-planar to the galaxy disk (inclination $i$\,=\,57.7{\degree}), the de-projected maximum velocity and outflow extent are $V_{\rm{out}}$\,$\sim$\,140\,{\kms} and $R_{\rm{out}}$\,$\sim$\,560\,pc, respectively. 
Using this values in Eq.\,\ref{eq:outflow}, we obtained a mass outflow rate of $\dot{M}_{\rm{mol,out}}$\,$\sim$\,0.1\,--\,1\,{\msunyr}. 

\paragraph{\textbf{Mrk\,915 (trailing arms scenario).}}
Considering only the CO emission from the component \texttt{c2} with negative LoS residuals (nearest two clumps, see Fig\,\ref{fig:bertola}), the mass of the outflowing clouds is $M_{\rm{mol,out}}$\,$\sim$\,$0.8$\,--\,$7$\,$\times$\,$10^6$\,{\msun} (from $S_{\nu} \Delta \vel_{\rm{CO(2-1),out}}$\,$=$\,184$\pm$4\,{\mjykms}), which is 4 percent of the total cold molecular mass. Therefore, using Eq.\,\ref{eq:outflow}, the cold molecular gas mass outflow rate is $\dot{M}_{\rm{mol,out}}$\,$\sim$\,0.09\,--\,0.7\,{\msunyr}, for $V_{\rm{out}}$\,$\sim$\,300\,{\kms} and $R_{\rm{out}}$\,$\sim$\,2.8\,kpc (de-projected values), where we assumed an outflow inclination co-planar with the disk ($i$\,=\,66.5{\degree}).

\paragraph{\textbf{MCG\,-01-24-012 (CND scenario).}}
The total CO(2-1) flux in the CND double-peak region of MCG\,-01-24-012 is $S_{\nu} \Delta \vel_{\rm{CO(2-1),out}}$\,$\sim$\,9.22$\pm$0.01\,{\jykms} including the flux of both components. This corresponds to a molecular gas mass of $\sim$\,$3$\,--\,$20$\,$\times$\,$10^7$\,{\msun}, representing  $\sim$\,30 percent of the total observed cold molecular mass. 
For a molecular outflow close to plane of the CND (inclination $i$\,=\,74{\degree} in our preferred model), the corresponding mass outflow rate is $\dot{M}_{\rm{mol,out}}$\,$\sim$\,0.4\,--\,3\,{\msunyr}. 
To obtain $\dot{M}_{\rm{mol,in}}$, we used Eq.\,\ref{eq:outflow}, with a de-projected outflow velocity and extent of $V_{\rm{out}}$\,$\sim$\,40\,{\kms} and $R_{\rm{out}}$\,$\sim$\,3\,kpc ($\sim$\,2{\arcsec}), respectively. The $V_{\rm{out}}$ corresponds to the mean absolute difference between the velocities of the components relative to our preferred disk model for the CND.

\section{Additional figures}\label{ap:additional_fig}

\begin{figure*}
	\includegraphics[width=1\linewidth]{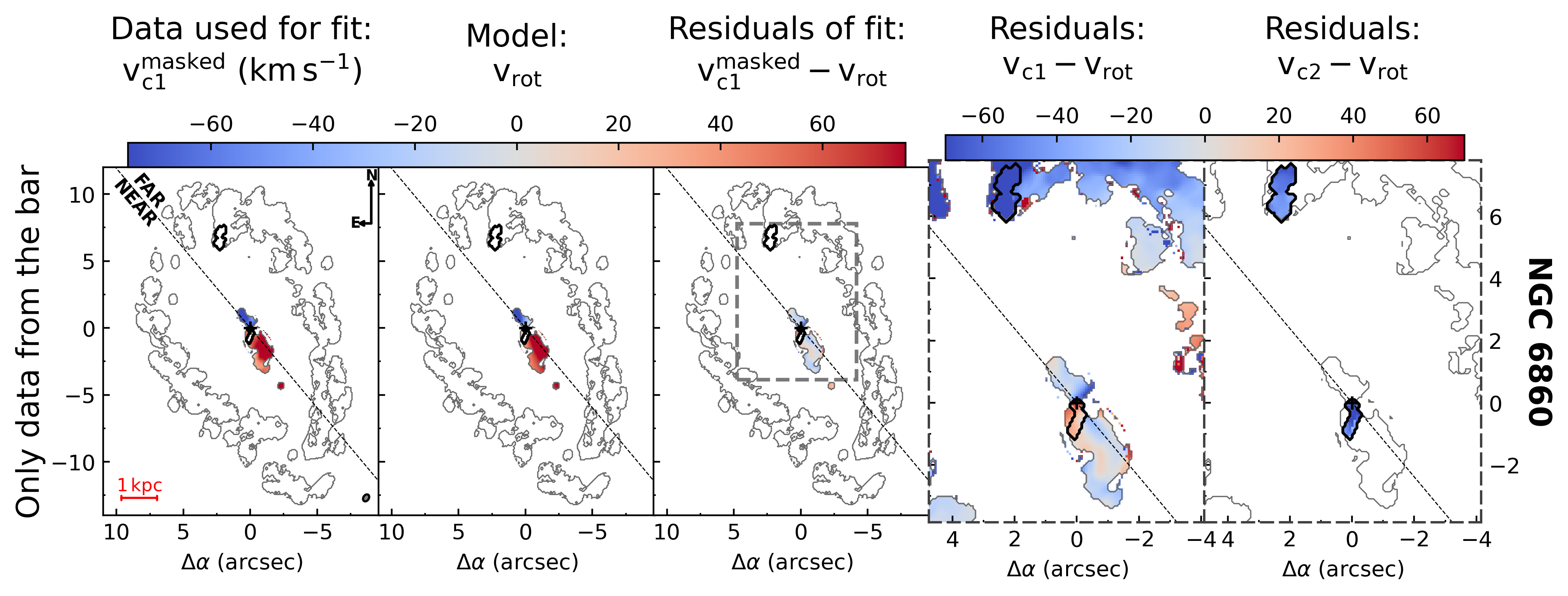}
    \caption{
    Similar to Fig.\,\ref{fig:bertola}, but for the fit done using only the CO data (from component \texttt{c1}) along the stellar bar in NGC\,6860. 
    }
    \label{fig:bertola-ngc860_onlybar}
\end{figure*}

\begin{figure}
	\includegraphics[width=1\linewidth]{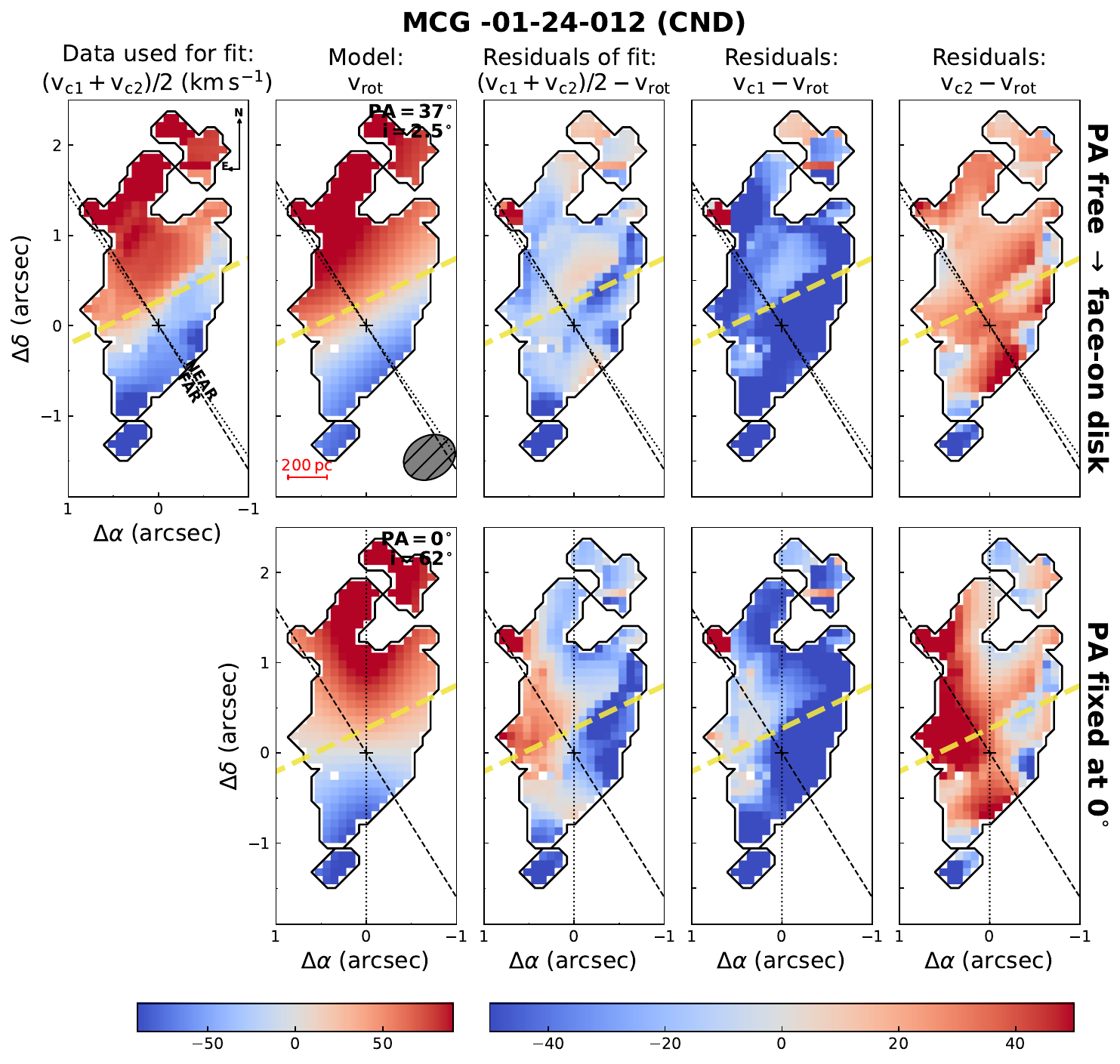}
    \caption{
    Similar to Fig.\,\ref{fig:bertola}, but for the fit done in the region with double-peak profiles of MCG\,-01-24-012 to model the CND rotation. 
    In the first row, the PA parameter was set free during the fit, which resulted in a face-on disk, while in the second row, we fixed PA at 0\degree, resulting in a more inclined disk.
    The data used to fit the disk model corresponds to the average between the LoS velocities of the components \texttt{c1} and \texttt{c2} in the double-peak region (black contours). 
    The first three columns show the LoS velocity data, the best-fit rotation model, and the residuals of the fitted data. 
    The last two columns correspond to the residuals of each component. 
    The dotted and dashed black lines are the kinematic major axes of the CND models and the global model (Fig.\,\ref{fig:bertola}).
    The orange dashed line indicates the position of the stellar bar. 
    }
    \label{fig:bertola-mcg-inner-vel}
\end{figure}

\end{appendix}

\end{document}